\documentclass[twocolumn]{aastex63}
\submitjournal{ApJ}

\shorttitle{2017 Outburst of H~1743--322}
\shortauthors{Sahu et al.}
\usepackage[flushleft]{threeparttable}
\usepackage{url}
\usepackage{graphicx}

\begin{document}

\title{2017 Outburst of H~1743--322: \textit{AstroSat} and \textit{Swift} View}

\correspondingauthor{Parijat Thakur}
\email{parijat@associates.iucaa.in, parijatthakur@yahoo.com}

\author{Pragati Sahu}
\affil{Department of Pure and Applied Physics, Guru Ghasidas Vishwavidyalaya (A Central University), Bilaspur (C. G.)- 495009, India}

\author{Swadesh Chand}
\affiliation{Inter-University Centre for Astronomy and Astrophysics, Post Bag 4, Ganeshkhind, Pune - 411007, India}

\author{Parijat Thakur}
\affiliation{Department of Pure and Applied Physics, Guru Ghasidas Vishwavidyalaya (A Central University), Bilaspur (C. G.)- 495009, India}

\author{G. C. Dewangan}
\affiliation{Inter-University Centre for Astronomy and Astrophysics, Post Bag 4, Ganeshkhind, Pune - 411007, India}

\author{V. K. Agrawal}
\affiliation{Space Astronomy Group, ISITE Campus, U. R. Rao Satellite Center, Outer Ring Road, Marathahalli, Bangalore - 560037, India}

\author{Prakash Tripathi}
\affiliation{State Forensic Science Laboratory, Raipur (C. G.)- 492001, India}

\author{Subhashish Das}
\affil{Department of Pure and Applied Physics, Guru Ghasidas Vishwavidyalaya (A Central University), Bilaspur (C. G.)- 495009, India}





\begin{abstract}
We perform a comprehensive timing and broadband spectral analysis using an AstroSat observation of the low-mass black hole X-ray binary H~1743--322 during 2017 outburst. Additionally, we use two Swift/XRT observations, one of which is simultaneous with AstroSat and the other taken three days earlier, for timing analysis. The hardness-intensity diagram indicates that the 2017 outburst was a failed one unlike the previous successful outburst in 2016. We detect type C quasi-periodic oscillation (QPO) in the simultaneous AstroSat and Swift/XRT observations at $\sim0.4$ Hz, whereas an upper harmonic is noticed at $\sim0.9$ Hz in the AstroSat data only. Although these features are found to be energy independent, we notice a shift of $\sim0.08$ Hz in the QPO frequency over the interval of three days. We also investigate the nature of variability in the two consecutive failed outbursts in 2017 and 2018. We detect soft time lags of $23.2\pm12.2$ ms and $140\pm80$ ms at the type C QPO frequencies in 2017 Astrosat and 2018 XMM-Newton data, respectively. The lag-energy spectra from both the outbursts suggest that the soft lags may be associated with the reflection features. The broadband spectral analysis indicates that the source was in the low/hard state during our AstroSat observation. Modeling of the disk and reflection continuum suggests the presence of a significantly truncated accretion disk by at least $27.4~r_{\rm{g}}$ from the ISCO when the source luminosity is $\sim1.6\%$ of the Eddington luminosity.
\end{abstract}


\section{Introduction} \label{sec:intro}

In black hole X-ray binaries (BHXRBs), a stellar mass black hole accretes material in the form of a disk from its normal companion star. The accretion disk, as it gets heated up to  $10^7$ K due to viscous forces between different layers in the inner regions, is an efficient emitter of X-rays. The broadband X-ray emission from BHXRBs consists of multicolor blackbody emission from an optically-thick and geometrically-thin accretion disk \citep{Shakura and Sunyaev1973}, a power-law component due to the Compton up-scattering of disk photons in a hot corona ($kT_e \sim 100$ keV), and a reflection component that arises due to the irradiation of coronal X-rays on to the disk \citep{Shakura and Sunyaev1973, Sunyaev and Titarchuk1980, Zdziarski et al.2004, Done et al.2007}. The most prominent features of the reflection spectrum are the ﬂuorescent iron $\mathrm{K}_\alpha$ line near $6.4$ keV and a reflection hump peaking at $\sim15-30$ keV band \citep{Fabian et al.2000}. However, the shape of the iron line may appear to be distorted due to the relativistic effects in the close proximity of the black hole. The study of the reflection spectrum can shed light on the accretion mechanisms at the close vicinity of the black hole, disk-corona geometry and inner disk radius  \citep{Fabian et al.1989, Ingram et al.2016}. Almost all the known low-mass BHXRBs are found to be transient in nature, and show sporadic outbursts due to abrupt changes in their mass accretion rate from the prolonged periods of quiescence. The luminosity of the black hole transients (BHTs) may change by a several orders of magnitude during an outburst which can last for several days or months \citep{Shidatsu et al.2014, Plant et al.2015}. For the successful or full outbursts, BHTs evolve through different spectral states following a q-shaped track in an anti-clock wise manner in the hardness-intensity diagram \citep[HID;][]{Belloni et al.2005}. At the beginning of such an outburst, the source stays in the low/hard state (LHS), where the energy spectrum is mainly dominated by the power-law component with photon index, $\Gamma\sim1.6$, and a high-energy cutoff around $100$ keV. This state is associated with a strong fractional rms variability of $\sim30\%$ \citep{Belloni et al.2005, Remillard and McClintock2006} and comparatively low luminosity. In addition to these, the accretion disk is usually thought to be truncated away from the inner most stable circular orbit \citep[ISCO;][]{McClintock et al.1995, McClintock et al.2003, Narayan and Yi1995, Narayan et al.1996, Motta2016}. Following the LHS, the source enters the intermediate states, namely, the hard-intermediate state (HIMS) and soft-intermediate state (SIMS), where the energy spectrum is contributed significantly by both the disk and power-law components. Moreover, moderate fractional rms variability of $\sim5-20\%$ \citep{Motta2016} and photon index $\Gamma\sim1.6-2.5$ are observed in the intermediate states \citep{Motta et al.2009, Belloni et al.2006}. Finally, the source moves to the high/soft state (HSS), which is strongly dominated by the emission from the disk and a weak power law component with $\Gamma\sim2.5-4$. The variability of the source in this state turns out to be very low (only a few percent), and the accretion disk is usually found to be extended close to the ISCO \citep{Motta et al.2009,  Motta2016}. During an outburst, timing properties of the source are also found to evolve over the different spectral states. Appearance of the low-frequency quasi-periodic oscillations (QPOs) in the power density spectra (PDS) of different spectral states are noticed roughly in the frequency range of $0.1-30$ Hz. These low-frequency QPOs are categorised into three types; type A, B and C \citep[see][]{Casella et al.2004, Casella Belloni and Stella2005, Remillard and McClintock2006}. The type C QPOs are observed usually in the LHS and HIMS, whereas type A and type B QPOs occasionally appear in the SIMS. On the other hand, weak QPOs are observed in the HSS \citep{Casella et al.2005, Belloni2010}. It is worth mentioning here that the BHTs do not always evolve through all the states in an outburst, as some exceptions have been found in which the sources remain in the LHS over the entire outburst periods and never enter the HSS. These kind of outbursts are termed as failed or hard-only outburst \citep{Capitanio et al.2009, Alabarta et al.2021}.

The BHTs show variability over a timescale of seconds to days, which can be associated with different emission mechanisms and connections between them. The nature of the variability may also be dependent on the change in source flux and energy due to different
ongoing physical processes. However, the exact origin of these variability processes are still not clear. Two prominent approaches to investigate the variability properties in BHTs are the study of rms-energy spectrum and time lag between different energy bands. The study of rms-energy spectrum helps to understand the origin of the energy dependent variability by distinguishing between the variable and constant emission spectral components. Another intriguing aspect to understand the nature of variability is the study of lag-frequency and lag-energy spectra. The measured time lags appear to be either positive or negative. The positive lag, often called as hard lag, indicates that the hard energy photons are lagging behind the soft ones. The hard lags are usually observed at low-frequency and thought to be originated due to the propagation of fluctuations in the mass accretion rate in the accretion disk \citep{Lyubarskii1997, Nowak et al.1999, Arevalo and Uttley2006, Uttley et al.2011, DeMarco et al.2013}. On the other hand, the soft or negative lags, where the soft photons follow the hard ones, appear comparatively at higher frequencies than the hard lags. These soft lags are interpreted as the time delay between the reflected emission from the disk and the primary X-ray emission. The study of the soft lags can provide important information about the geometry of the innermost region in the close proximity of the black hole \citep{Uttley et al.2014, Kara2019}. 

\begin{table*}
	\caption{Observation details of 2017 and 2018 outbursts of H 1743--322}

   \centering
   \renewcommand{\arraystretch}{1.2}
   \begin{tabular}{lcccc}
      \hline
      \hline
  
    Obs  & Instrument & Obs ID & Obs Date &  Eff. Exp.(ks) \\
     \hline
     & & \texttt{2017 Outburst: AstroSat} & & \\
     \hline
     data--1 & AstroSat/SXT & $ G07\_039T01\_9000001444$ & $2017$ August $08$  &   12.8  \\
        &      AstroSat/LAXPC   & ... &   ... &  20.22 \\
        \hline
        & & \texttt{2017 Outburst: Swift} & & \\
        \hline
   data--2 & Swift/XRT & $00031441074$ &  $2017$ August $05$ & 1.3 \\
    data--3 & Swift/XRT & $00031441075$ & $2017$ August $08$ & 1.6 \\

      \hline
      & & \texttt{2018 Outburst: XMM-Newton} & & \\
      \hline
   data--4& XMM-Newton/EPIC-pn & $0783540401$ & 2018 September 26 & 127.2\\
     \hline
   \end{tabular}
\end{table*}     

Being discovered by the Ariel--V satellite in 1977 \citep{Kaluzienski1977}, the low-mass BHT H~1743--322 has showed frequent outbursts. \citet{Shidatsu et al.2012, Shidatsu et al.2014} suggested an apparent interval of $\sim200$ days for such outbursts of this source. Using the radio and X-ray observations of its brightest outburst in 2003, \citet{Steiner et al.2012} reported the source distance, inclination angle and spin to be $8.5\pm0.8$ kpc, $75^\circ\pm3^\circ$ and $0.2\pm0.3$, respectively. Evidence of high inclination angle has also been reported by \citet{Homan et al.2005} and \cite{MunozDarias et al.2013} for this source. Following the 2003 outburst, H~1743--322 went through many frequent outbursts until 2016. However, the nature of the outbursts for H~1743-322 never remained the same with time, and exceptions were found when the source stayed in the LHS for the entire outburst periods unlike the successful outburst. These kind of failed or hard-only outburst was first reported by \citet{Capitanio et al.2009} for the 2008 outburst. Later on, several other failed outbursts were also seen for this source \citep[e.g.,][]{Stiele and Yu 2016, Tetarenko et al.2016}.  Apart from these failed outbursts, four successful outbursts were seen in 2009, 2010, 2013 and 2016 \citep{Motta et al.2010, Chand et al.2020, Alabarta et al.2021}. Extensive study of these outbursts has resulted in a detection of QPOs from mHz to high-frequency range \citep{Homan et al.2005, Remillard et al.2006, Sriram et al.2009, Altamirano and Strohmayer2012, Stiele and Yu 2016, Chand et al.2020, Chand et al.2021}. Although the exact reason responsible for the failed outburst is not yet clear, \citet{Capitanio et al.2009} suggested that the lack of soft-state transitions is probably connected to a premature decrease of the mass accretion rate. \citet{Alabarta et al.2021} carried out a study to differentiate between failed and successful outbursts using a sample of low-mass BHTs and suggested that failed and successful outbursts may be different manifestations of one unique accretion-related process. However, different nature in the variability processes were seen between these two types of outbursts. In both 2008 and 2014 failed outbursts of H~1743--322, \citet{DeMarco and Ponti2016} reported the presence of reverberation lag of $\sim60$ ms, which is the light-crossing time delay between the primary and the reprocessed emission. On the other hand, presence of hard lag, originating due to the propagation of fluctuations in the mass accretion rate from the outer part of the disk to the inner part, was found in the system during the 2016 successful outburst by \citet{Chand et al.2020}. However, no change in the Eddington luminosity between the three outbursts may indicate that the different nature of variability was not triggered by the source luminosity rather may be by some other physical process \citep{Chand et al.2020}. Following the 2016 successful outburst, H~1743--322 showed two other consecutive failed outbursts in 2017 \citep{Jin et al.2017} and 2018 \citep{Stiele and Kong2021, Wang et al.2022}. 

As \citet{Husain2023} considered only single AstroSat observation of 2017 outburst to study the nature of PDS in the $3-15$ keV band, their work does not provide the information regarding the energy and time dependent properties of the QPO. Moreover, they do not perform the reflection analysis of the time-averaged X-ray energy spectrum. This  motivates us to further carry out a comprehensive timing and broadband spectral analysis  during the 2017 outburst. In this paper, we derive HID for the 2017 outburst and compared it with that of the previous successful outburst in 2016. We detect the presence of type C QPO along with upper harmonic, and investigate their energy dependent behavior in three different energy bands. Using AstroSat and Swift observations, we find that the QPO frequency changes over an interval of three days. The broadband spectral investigation reveals the presence of an iron line and reflection hump in the time averaged energy spectrum. We employ two different combinations of model components including the relativistic reflection model \texttt{relxillCp} \citep{Dauser2014, Garcia2014} for spectral analysis. In order to perform the comparative time lags between the two consecutive failed outbursts, we have carried out time lag analysis using AstroSat and XMM-Newton observations  taken during 2017 and 2018 outbursts, respectively. The remainder of the paper is organized as follows. We present the observations and data reduction in Section 2, and Section 3 contains the analysis and results of our timing and broadband spectral studies. Finally, Section 4 is devoted to discussion and concluding remarks. 

\begin{figure*}
\centering
\includegraphics[scale=0.5]{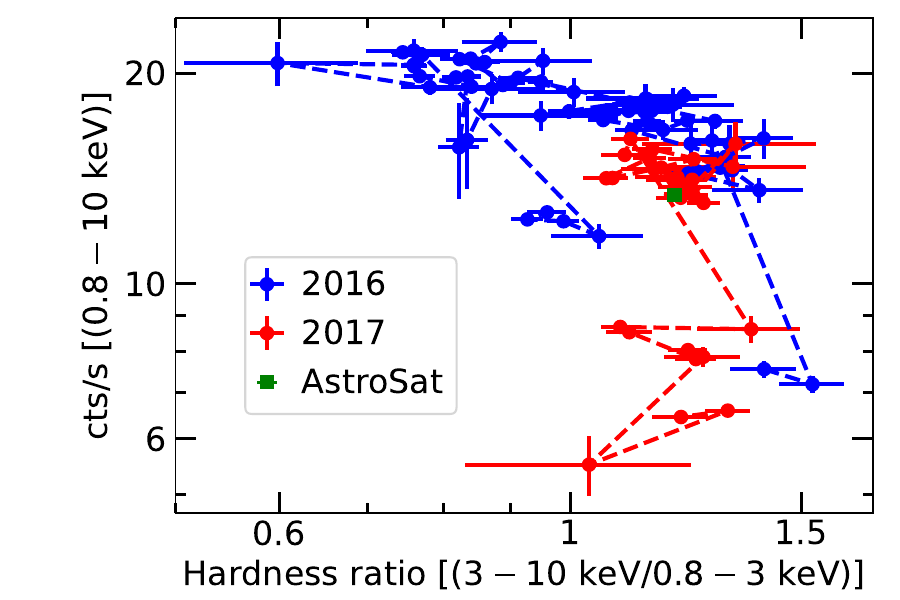}
\includegraphics[scale=0.5]{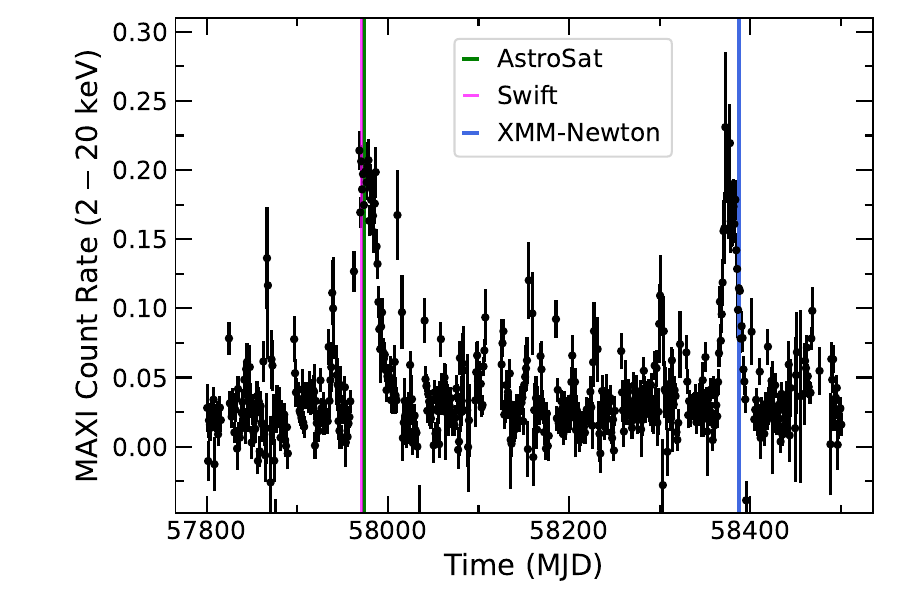}
\caption{Left panel: Hardness intensity diagram derived using Swift/XRT windowed timing mode data taken during 2016 and 2017 outbursts. The blue points represent 2016 outburst, whereas the red points denote 2017 failed outburst. The hardness is derived between $0.8-3$ keV and $3-10$ keV bands. The green point denotes the start time of AstroSat observation. Right panel: MAXI longterm lightcurve in the 2-20 keV band covering 2017 and 2018 outbursts of H~1743--322. The vertical lines indicate the positions of the Swift (magenta) and AstroSat (green) observations during 2017 outburst, and XMM-Newton (blue) observation during 2018 outburst.}
\end{figure*}

\section{Observation and Data Reduction} \label{sec:style}

We use publicly available one AstroSat and two Swift/XRT datasets  acquired during  the 2017 outburst of H~1743--322. In case of Swift/XRT, we consider only those data sets with exposure time more than 1 ks. While we utilize the AstroSat data for both the timing and spectral analysis, we only performed timing analysis of Swift/XRT data due to the poor signal-to-noise (S/N) and short exposures.  These observations were taken when the source was at the peak of its outburst and are marked on the MAXI light curve, shown in the right panel of Figure~1. For a comparative study of time lags between the 2017 and 2018 outbursts, we also consider XMM-Newton/EPIC-pn observation taken during the 2018 outburst along with the AstroSat observation of 2017 outburst. This XMM-Newton observation was taken during the decaying phase of 2018 outburst and is also marked in MAXI light curve shown in right panel of Figure~1. The details of data analyzed here are listed in Table~1. To derive the HID, we have used all Swift/XRT data taken in windowed timing mode during the 2017 outburst of the source and derived the count rates in $0.8-3$ keV, $3-10$ keV and $0.8-10$ keV bands. The HID is shown in the left panel of Figure 1, where the hardness ratio is derived by dividing the count rate in the $3-10$ keV band with that of the $0.8-3$ keV band. Apart from this, we have also derived the HID using the Swift/XRT observations of the 2016 successful outburst 
in a similar way for the sake of comparison with the 2017 failed outburst (see Figure~1). It is clear from both the HIDs that the source remains in the hard state during the entire period of 2017 outburst and never entered the soft state. 

\begin{figure*}
\plottwo{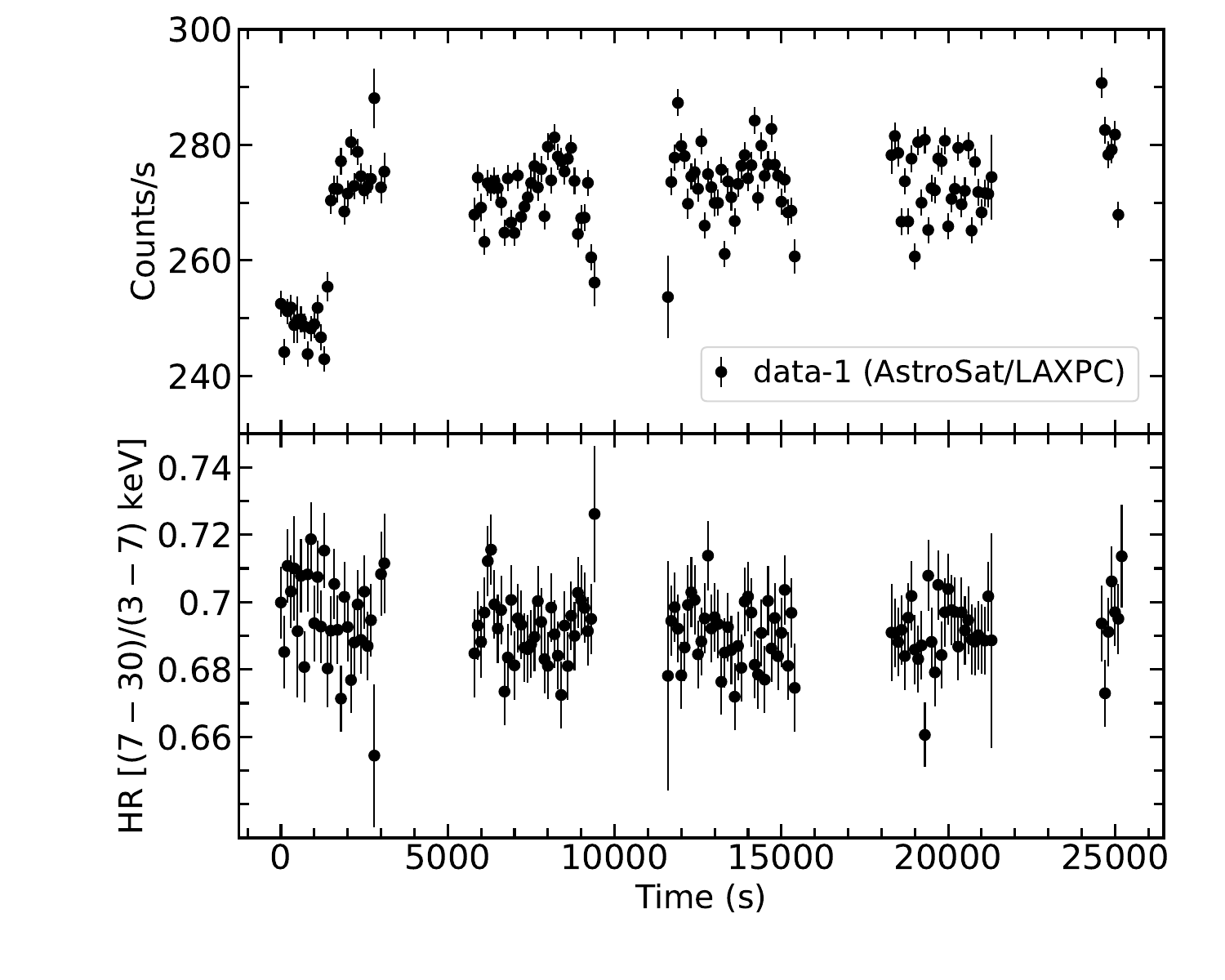}{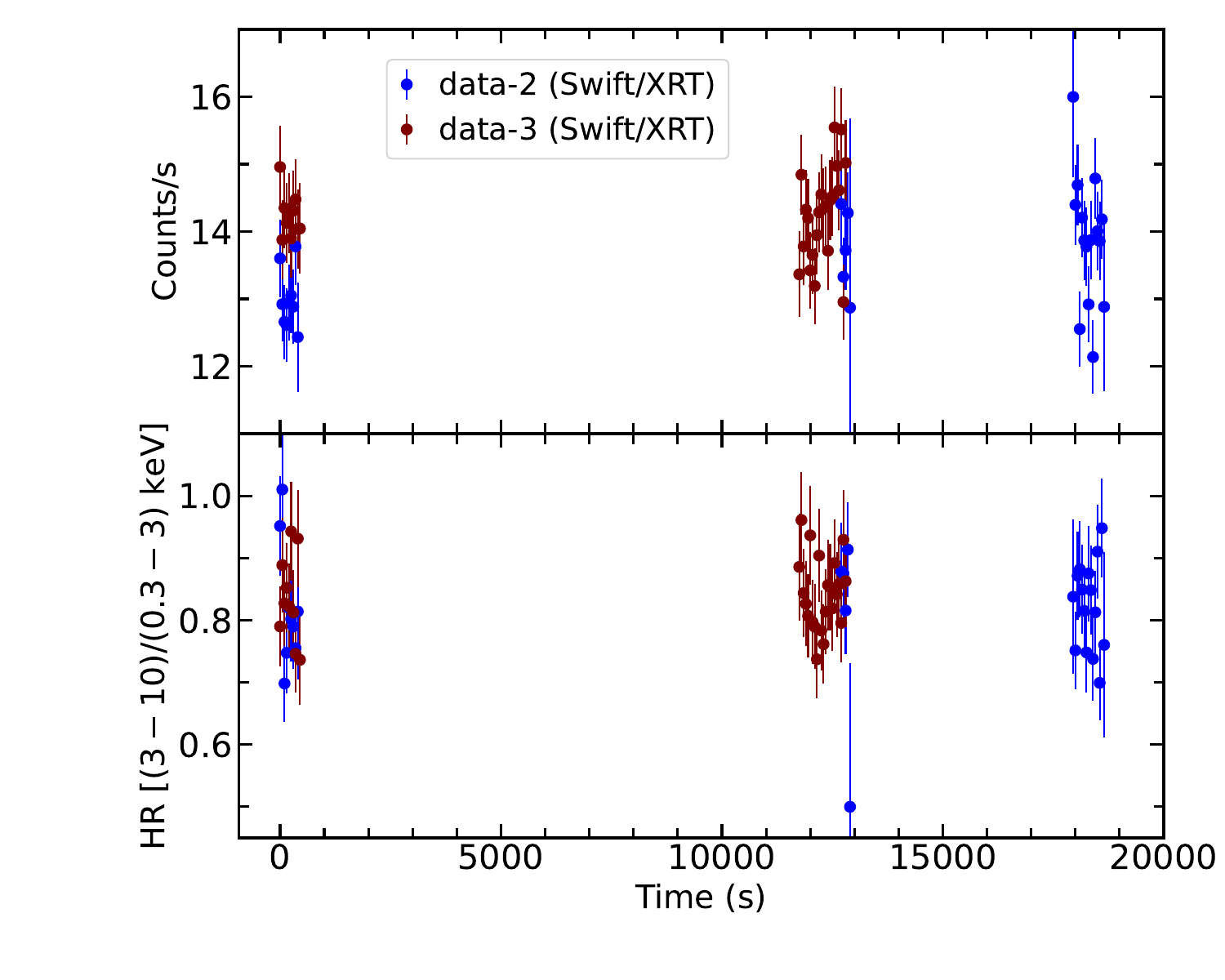}
\caption{Left: Background subtracted light curve with the time bin size of 100 s derived in the $3-80$ keV band using data--1 taken by LAXPC20 (upper panel) and corresponding HR derived between $3-7$ and $7-30$ keV bands (lower panel). Right: Background subtracted light curve with the time bin size of 50 s in the $0.3-10$ keV band using data--2 (blue filled circle) and data--3 (maroon filled circle) taken  by Swift/XRT (upper panel) and corresponding HR derived between $0.3-3$ keV and $3-10$ keV bands (lower panel).}
\end{figure*}

\subsection{AstroSat/SXT}
     
Soft X-Ray Telescope \citep[SXT;][]{Singh et al.2016, Singh et al.2017} onboard AstroSat is useful at the low energy band of $0.3-8.0$ keV. We downloaded level 2 SXT event files from ISSDC 
website\footnote{\url{https://webapps.issdc.gov.in/astro_archive/archive/Home.jsp}} and used the Julia event merger tool\footnote{\url{https://www.tifr.res.in/~astrosat_sxt/dataanalysis.html}} 
to create a merged clean event file. A circular region of $15\arcmin$ centered at the source PSF was used to extract the source spectrum with \texttt{XSELECT} (V2.6d) tool. The source count rate 
is found to be $5.44\pm0.02$ cts/s. This is much lesser than the threshold value of the count rate ($\gtrsim 40 $ cts/s) for the pile up 
in the SXT PC mode observation. Since the large PSF (FWHM$\sim2\arcmin$) of the SXT with the broad wings extending to $\sim15\arcmin$ leaves no source-free area in the CCD detector, we used 
the SXT blank sky spectral file `SkyBkg$\_$comb$\_$EL3p5$\_$Cl$\_$Rd16p0$\_$v01.pha' for the background spectrum as recommended by the instrument team. For the response matrix file (RMF), we have used `sxt$\_$pc$\_$mat$\_$g0to12.rmf' provided by the SXT Payload Operation Center (POC) team. Using the \texttt{sxtARFModule} tool \footnote{\url{https://www.tifr.res.in/~astrosat_sxt/dataanalysis.html}}, we have created the SXT off-axis auxiliary response file (ARF) in accordance with the location of the source on the CCD from the on-axis ARF `sxt\_pc\_excl00\_v04\_20190608\_corr1kev\_sharp\_chng.arf' provided by the SXT POC  team to perform correction for vignetting effect. Since the current version of SXTPIPELINE does not take into account the shift in the gain of onboard SXT instrument, we have corrected the SXT spectrum for this gain shift using the gain command in XSPEC (V.12.10)\footnote{\url{https://heasarc.gsfc.nasa.gov/xanadu/xspec/}} within HEASoft. For this, the slope was fixed at unity and the offset 
was taken to be variable.

\subsection{AstroSat/LAXPC}

We have used the data acquired by the Large Area X-Ray Proportional Counter \citep[LAXPC;][]{Yadav et al.2016a, Yadav et al.2016b, Agrawal et al.2017, Antia et al.2017}. By employing the LAXPC software\footnote{\url{http://astrosat-ssc.iucaa.in/?q=laxpcData}} (Laxpcsoft), the level 1 data of the source was processed to obtain the level 2 data. Further, the lightcurves and energy spectrum are produced using the standard tasks available within Laxpcsoft\footnote{\url{https://www.tifr.res.in/~astrosat_laxpc/LaxpcSoft.html}}.
For the spectral and timing analysis, only the data from LAXPC20 detector is considered as LAXPC30 detector is no longer functioning and LAXPC10 is working at a low gain.

\subsection{Swift/XRT}
We have considered Swift/XRT observations taken in the windowded timing (WT) mode and created level 2 event file from the level 1 data using the \texttt{xrtpipeline} task. To extract the source lightcurve in the $0.3-10$ keV band, we used \texttt{XSELECT V.2.4} and considered a circular region of $30''$ centered at the position of the source. On the other hand, the background light curve was extracted in the same energy band using a circular region of $60''$ taken away from the source. The source lightcurve is corrected for the background using the \texttt{lcmath} task, where the scaling factor was chosen accordingly with the sizes of the source and background regions.

\subsection{XMM-Newton}
We have used the data taken by the XMM-Newton with European Photon Imaging Camera (EPIC-pn) in timing mode with a thick filter. 
To filter and extract the EPIC-pn event file, we used the Science Analysis System (SAS V.21) along with latest calibration files. We noticed the presence of background flaring in the lightcurve derived in the $10-12$ keV band, which was eliminated by creating a good time interval (GTI) with the condition of count rate $\le1.2 \mathrm{~s^{-1}}$. This GTI file was then employed to create the cleaned event file. Following this, we adopted the criteria outlined in \citet{Stiele and Kong2021} to extract source and background regions, as well as to mitigate pileup. The lightcurves were also corrected for the background using the task \texttt{epiclccorr}, embedded within SAS.

\begin{figure*}
\gridline{
\fig{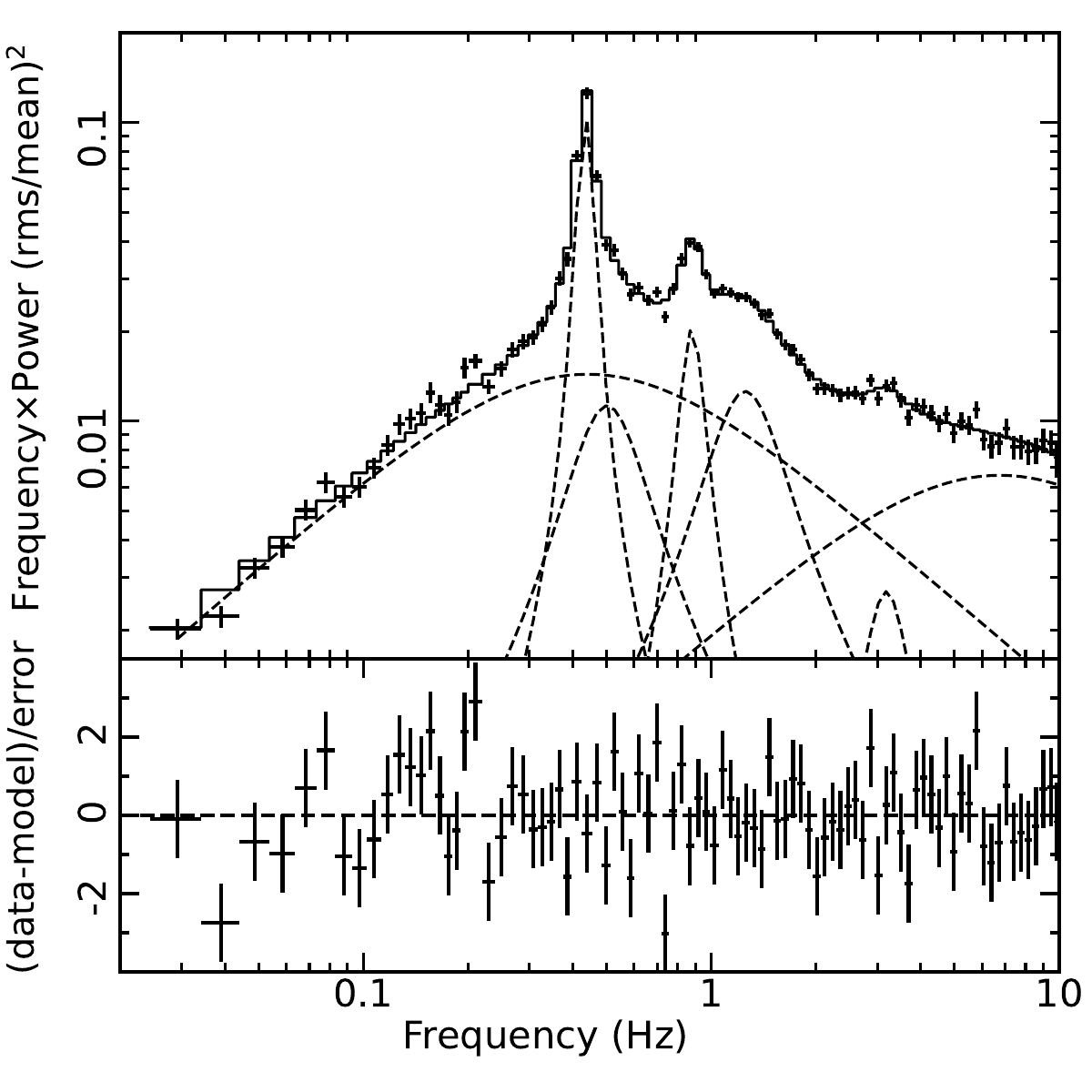}{0.315\textwidth}{}
\fig{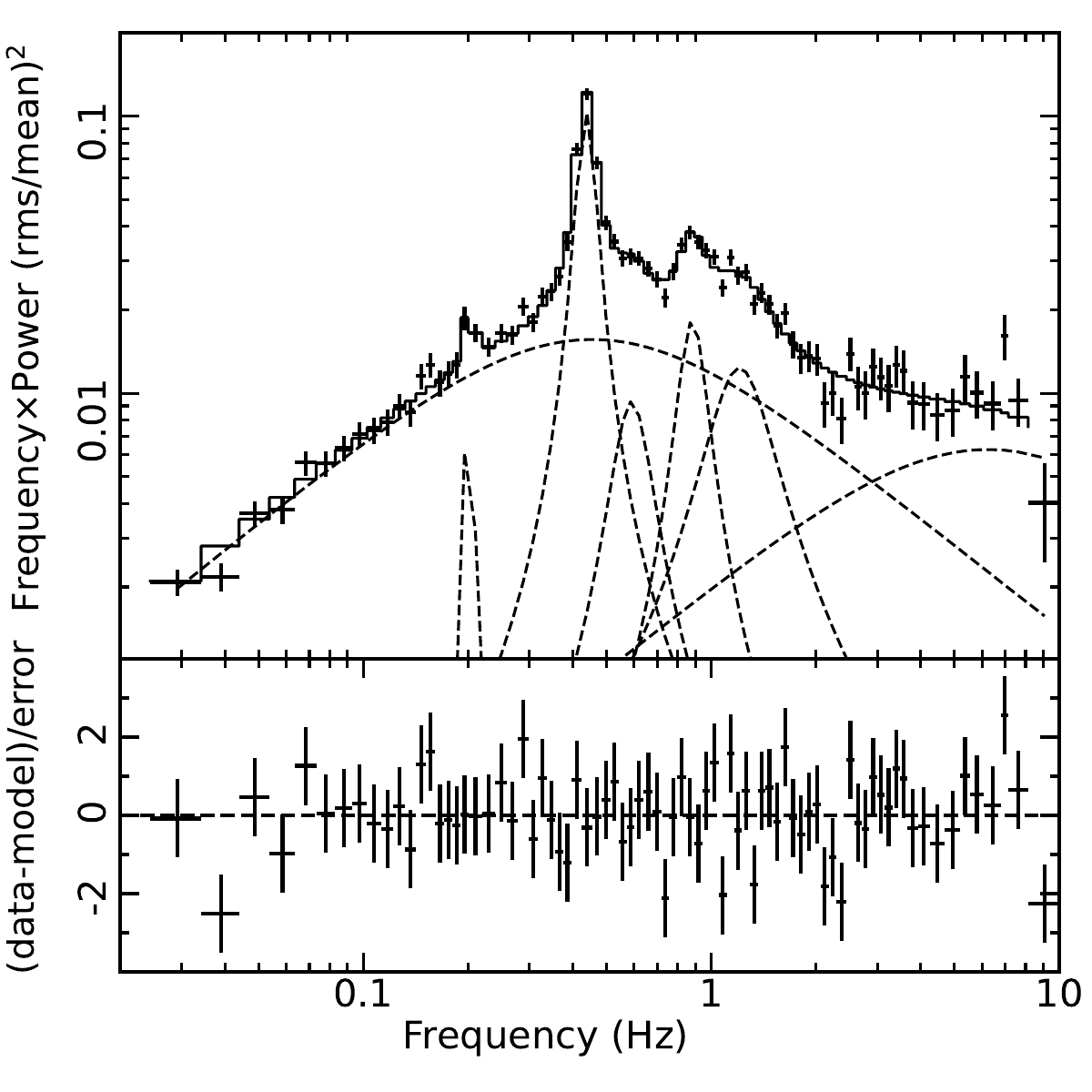}{0.315\textwidth}{}
\fig{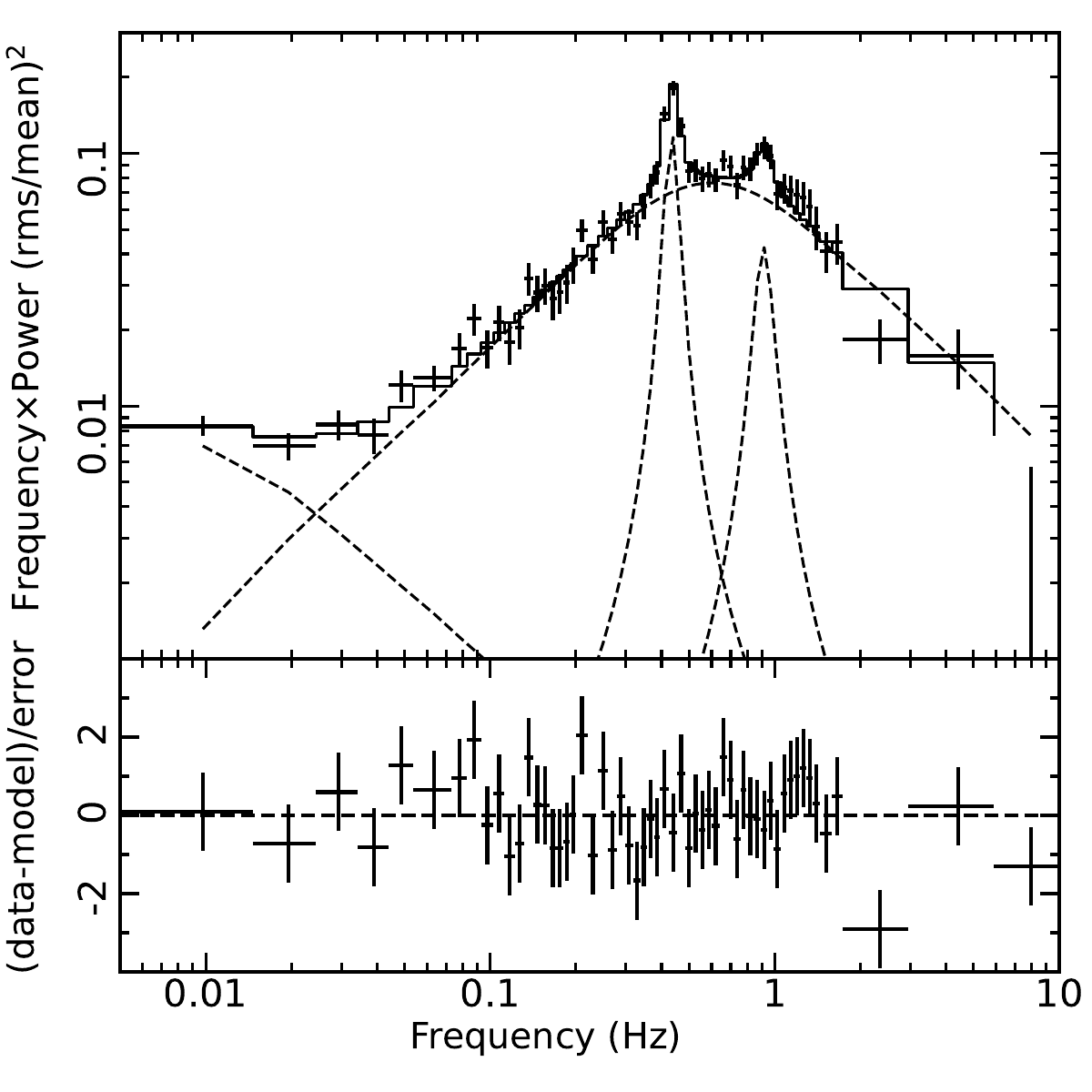}{0.315\textwidth}{}} 
\caption{Power density spectra in the $3-15$ keV (left panel), $15-30$ keV  (middle panel) and $30-50$ keV (right panel) bands derived from the data--1 taken by AstroSat/LAXPC20. Solid lines indicate the best-fit model and the dotted lines show the individual model components.\label{fig:f3}}
\end{figure*}

\section{Timing Analysis and Results} \label{sec:floats}

We derived the source and background lightcurves
from  data--1 taken by AstroSat/LAXPC20 with a time bin size of 100 s in the full energy band of $3-80$ keV, and then corrected the source lightcurve for the background using the lcmath task within FTOOLS. The upper left panel of Figure 2 exhibits the background subtracted lightcurve for the full exposure time of the LAXPC20 data. The source intensity appears to be constant throughout the observation period with an average count rate of $\sim275$ cts/s. However, the slight drop in the count rate at the beginning of the lightcurve is caused by the angular offset in pointing the source during the start of the observation. As shown in the bottom left panel of Figure 2, the hardness ratio (HR), derived between $3-7$ and $7-30$ keV bands, does not exhibit any variation, indicating no spectral evolution of the source during this particular observation. 
We also derived background subtracted light curves with the time bin size of 50 s in the $0.3-10$ keV band and HR between $0.3-3$ keV and $3-10$ keV bands using data--2 and data--3 taken by Swift/XRT, shown in the upper and bottom right panels of Figure 2, respectively. As similar to data--1, no variation either in the count rates or HR are noticed over the full exposure time for data--2 and data--3.

\subsection{Power Density Spectra}
For the timing analysis, we used \texttt{XSPEC} and quoted all the error bars at $90\%$ confidence level. We derived PDSs at three different energy bands of $3-15$, $15-30$, and $30-50$ keV from data--1 taken by AstroSat/LAXPC20 using \texttt{laxpc\_find\_freq}\footnote{\url{http://astrosat-ssc.iucaa.in/laxpcData}} task considering the Nyquist frequency of 10 Hz. The data above 50 keV were excluded due to poor signal to noise ratio and large background. We considered the intervals of 2048 bins and derived the rms-normalized PDS for each interval by dividing the Leahy normalized power spectra with an average rate \citep{Leahy et al.1983, Belloni and Hasinger1990}. The final PDS was then derived by averaging all the rms-normalized PDSs for different intervals while propagating the error bars accordingly and rebinning the PDSs geometrically in frequency space by a factor of 1.05. Further, the dead time-corrected Poisson noise level was subtracted from each PDS derived in the above mentioned bands \citep{Zhang et al.1995, Yadav et al.2016b, Agrawal et al.2018, Chand et al.2022}. As the true source rms can be affected by LAXPC dead time of 43 $\mu$s, its effect was corrected using the formula rms$_{\rm{in}}=\rm{rms_{det}} (1+\tau_d r_{\rm{det}}/N)$ prescribed by \citet{Bachetti et al.2015}. Here, rms$_{\rm{in}}$ is the dead-time corrected rms, rms$_{\rm{det}}$ is the instrument detected rms, $\tau_d$ is the dead-time, $r_{\rm{det}}$ is the detected count rate and $N$ is the number of proportional counter units \citep{van der klis1988, Zhang et al.1995, Sreehari et al.2020}. Apart from this, PDSs were also corrected for the corresponding background count rates \citep{Yadav et al.2016b, Rawat et al.2019}.  
 
The PDSs in the $3-15$ and $15-30$ keV bands were fitted with seven Lorentzians, whereas the PDS in the $30-50$ keV band required only four Lorentzians. All the best-fit temporal parameters are listed in Table~2 and the fitted PDSs in all three bands are shown in Figure~3. We detected the presence of QPO and its upper-harmonic in the PDSs derived for the above mentioned three energy bands  at $\sim0.4$ Hz and $\sim0.9$ Hz, respectively. This indicates the energy independent nature of the QPO and its upper-harmonic. The detection significance of the QPO (upper-harmonic) are found to be  $\sim13\sigma$($\sim8.3\sigma$), $\sim11\sigma$($\sim3.8\sigma$) and $\sim8\sigma$($\sim4.9\sigma$) in the $3-15$, $15-30$ and $30-50$ keV bands, respectively. The quality-factor ($Q$=$\nu_{\rm{centroid}}$/FWHM; FWHM being the full width at half maximum) of QPOs remain the same within errors in all the three bands, whereas the fractional rms variability of the QPO slightly increases in the $30-50$ keV band with respect to the other two bands. In a similar way, the $Q$-factor and fractional rms variability of the upper harmonic remain almost same within error over the three energy bands.

\vspace{-5 cm}

\begin{figure*}
\plottwo{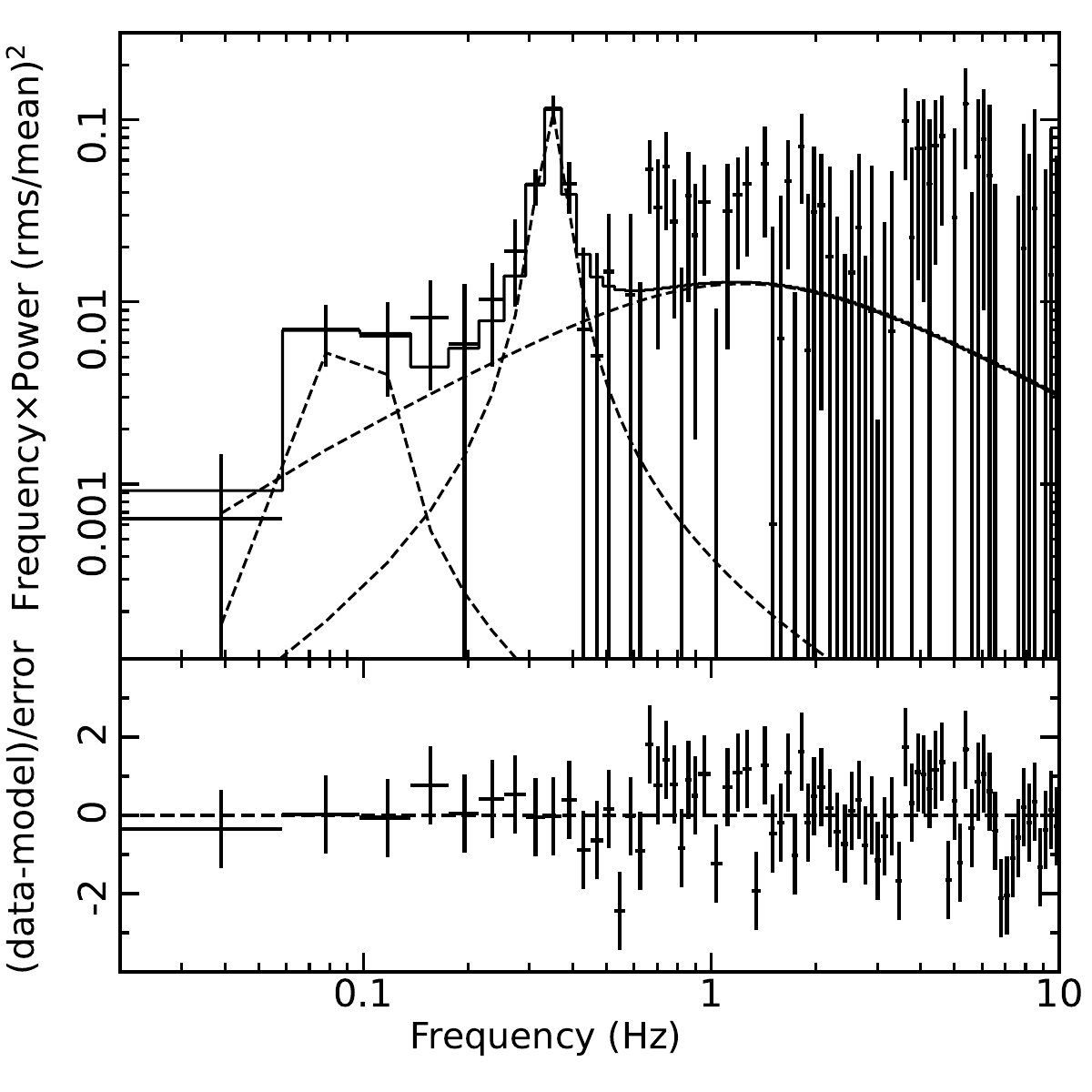}{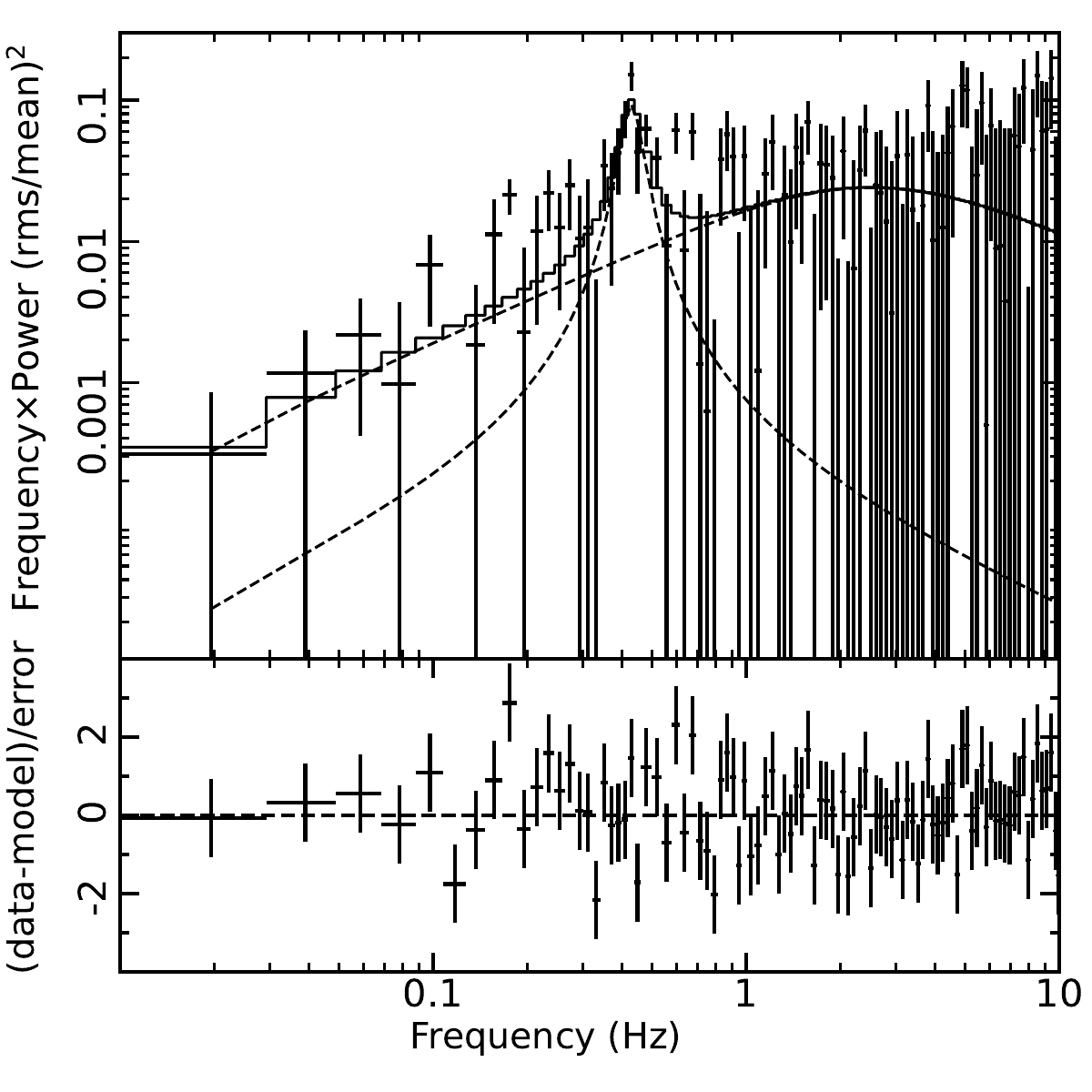}
\caption{Power density spectra derived from data--2 (left panel) and data--3 (right panel) using SWIFT/XRT in the $0.3-10$ keV  band. Solid lines indicate the best-fit model and dotted lines show the individual model components.}
\end{figure*}

\vspace{5cm}
\begin{table*}[!ht]
	\caption{Best-fit QPOs and upper-harmonic parameters obtained from the PDSs of 2017 outburst observations}
   \centering

   \setlength{\tabcolsep}{15pt}
   \renewcommand{\arraystretch}{1.2}
  \begin{tabular}{lccccc}
      \hline
      \hline

      &   \multicolumn{3}{c}{AstroSat/LAXPC}   &     \multicolumn{2}{c}{Swift/XRT} \\
    
    &   \multicolumn{3}{c}{data--1} &  {data--2}    & {data--3}\\
    
     &   \multicolumn{3}{c}{(08-Aug-2017)} &  {(05-Aug-2017)}    & {(08-Aug-2017)}\\
      \hline 
         
  Parameter & {$3-15$ keV} &{$15-30$ keV} & {$30-50$ keV} & {$0.3-10$ keV} & {$0.3-10$ keV} \\
         \hline 
   
       $\nu^{1}_{\rm{qpo}}$(Hz) & $0.43\pm0.002$ &  $ 0.44\pm0.002$  & $0.433\pm0.005$& $0.35\pm0.01$   &  $ 0.43^{+0.02}_{-0.01}$ \\
     
     $\rm{FWHM_{qpo}^2}$(Hz) & 0.04$\pm$0.006 & 0.05$^{+0.007}_{-0.006}$ & 0.057$\pm0.01$ &  0.04$\pm$0.02 & 0.07$^{+0.10}_{-0.05}$\\
     $Q^{3}_{\rm{qpo}}$ & 10.7$^{+1.9}_{-1.5}$ & 8.8$\pm1.0$ & 7.6$^{+2.3}_{-1.6}$ & $8.7^{+8.2}_{-2.7}$&  $6.1^{+14.9}_{-3.5}$ \\
     $\rm{rms_{qpo}^4}$[\%] & 13.2 $\pm0.9$ & 14.2$\pm0.6$ & 16.7$\pm1.7$& 16$^{+3}_{-2}$  &  15$\pm4$\\
     
     $\nu^{5}_{\rm{har}}$(Hz) & $0.876\pm0.008$ & $0.877\pm0.018$ & $0.91\pm0.03$ &..&..\\
     $\rm{FWHM_{har}^6}$(Hz) & 0.14$\pm0.03$ & 0.169$_{-0.062}^{+0.073}$ & $0.14^f$ &..&..\\
     $Q^{7}_{\rm{har}}$ & 6.2$^{+1.6}_{-1.0}$ & 5.2 $^{+2.9}_{-1.5}$ & 6.5$\pm0.2$&..&..\\
     $\rm{rms_{har}^{8}}$[\%] & 7.3 $\pm 0.7$ & 7.4$^{+1.5}_{-1.8}$ & 10.2$^{+1.6}_{-1.9}$&..&..\\
     
     $\chi^2$/dof &$106.38/71$ & $77.93/60$ & $49.2/47$ &$72.7/69$ & $105/92$\\   
         \hline 
   \end{tabular}
  \begin{tablenotes} 
 \item \textbf{Notes.} $^1$QPO frequency, $^2$Width of the QPO, $^3$Quality factor of the QPO, $^4$Fractional rms amplitude of the QPO, $^5$Upper-harmonic frequency, $^6$Width of the upper-harmonic,  $^7$Quality factor of the upper-harmonic,  $^8$Fractional rms variability of the upper-harmonic, f indicates the fixed parameter.
  \end{tablenotes}   
   
\end{table*}

For the Swift/XRT observations (data--2 and 3), we used the \texttt{POWSPEC} task within FTOOLS to extract PDSs from the lightcurves binned with 0.05 s in the $0.3-10$ keV band. Thus, the Nyquist frequency (10 Hz) remains same as in data--1. The PDS extracted from data--2 was well described with three Lorentzian components, whereas the PDS obtained from data--3 required only two Lorentzian components. Here, we have also detected the presence of QPOs at $\sim0.35$ (detection signification $\sim5.3\sigma$) and $\sim0.43$ Hz (detection significance $\sim3.1\sigma$) in the PDSs derived from data--2 and data--3, respectively. However, unlike data--1, we could not detect the presence of upper-harmonic in the PDSs obtained from these two data, which may be due to the poor  (S/N) ratio of the data. All the best-fit temporal parameters extracted from the PDSs of data--2 and data--3 are given in Table 2, and the best-fit PDSs are shown in Figure~4. As can be seen in this Table, the $Q$-factors and fractional rms variability for the Swift/XRT observations remain consistent and align well with the AstroSat results. Apart from this, the QPO frequencies appear to be the same for data--1 and data--3 taken by AstroSat/LAXPC20 and Swift/XRT on August 08, 2017, respectively. However, a clear shift in the centroid frequency of the QPO by $0.08\pm0.02$ Hz has been noticed between data--2 and data--3 taken over the interval of three days.

\subsection{Fractional RMS Variability Spectra}
Study of evolution of the fractional rms variability as a function of energy is important to understand the origin of the energy-dependent variability and the responsible spectral component behind this variability. In this regard, we first probed the fractional rms amplitude variability as a function of energy using the data--1 taken by AstroSat/LAXPC20. For this, we derived PDSs at $3-5$, $5-7$, $7-10$, $10-15$, $15-20$, $20-25$, and $25-30$ keV bands and calculated the fractional rms amplitude variability in the frequency range of 0.1-10 Hz for each PDS. The variation in fractional rms amplitude as a function of energy for data--1 is shown in the left panel of Figure 5. We notice that the fractional rms variability amplitude remains almost stable up to $\sim10$ keV and then a rising trend has been observed at higher energies, mainly dominated by the powerlaw and reflection components. However, we note here that the uncertainties in the fractional rms variability amplitude at the last three energy bands are very large. In addition to this, we have also studied the characteristics of the fractional rms amplitude variability with energy using data--2 and data--3 taken by Swift/XRT. In order to calculate the fractional rms energy spectra, we considered six different energy bands of $0.3-1$, $1-2$, $2-4$, $4-6$, $6-8$ and $8-10$ keV, and derived the fractional rms amplitude variability from the PDSs of each energy band in the same frequency range as used for data-1 acquired by AstroSat/LAXPC20 data. The low energy coverage of Swift/XRT up to $0.3$ keV has enabled us to search for the variability in the soft energy bands, mainly dominated by the emission from the accretion disk. The fractional rms amplitude variability as a function of energy for data--2 and data--3 are depicted in the right panel of Figure 5. Although the uncertainties are very large due to poor S/N ratio of the Swift/XRT observations, the fractional rms variability amplitude remains stable except for increasing trends at $\lesssim 1$ keV and
$\gtrsim 6$ keV regions.

We also probed the fractional rms as a function of photon energy using the XMM-Newton/EPIC-pn observation, taken during the decay
phase of 2018 failed outburst (see Figure~1). We calculated the fractional rms variability ampitude from the PDSs in the $0.6-1$, $1-2$, $2-4$, $4-6$, $6-8$ and $8-10$ keV bands
for the same frequency mentioned above for AstroSat/LAXPC and Swift/XRT observations. The rms-energy spectrum derived from the
XMM-Newton/EPIC-pn observation during 2018 outburst is shown in Figure~6. We notice that the fractional rms variability is higher at the lower energies ($\lesssim 1$ keV) and then
becomes flat up to $\approx 6$ keV. Following this, the fractional rms shows a drop in the $6-8$ keV band before exhibiting an increasing trend at higher 
energies ($>9$ keV). This drop in the fractional rms amplitude is most probably due to the contribution of the iron emission line originating at a large distance from the BH \citep{Stiele and Kong2021}.

\subsection{Frequency-dependent Lag and Lag-Energy Spectra}

To understand the physical mechanism responsible for the variability, occurring on various time-scale, we have derived time lags at the QPO frequency from data--1 using the LAXPC subroutine \texttt{laxpc\_find\_freq}\footnote{\url{http://astrosat-ssc.iucaa.in/laxpcData}}, which follows the 
cross-spectral analysis method as prescribed by \citet{Vaughan and Nowak1997} and \citet{Nowak et al.1999}. To calculate the error on the time lags, the technique given in \citet{Nowak et al.1999} was employed. In order to investigate the frequency-dependent time lags, we extracted two light curves in the $3-5$ and $9-13$ keV energy bands, which are usually dominated by the disk and powerlaw emissions, respectively. We then divided each light curve into 342 segments, each having the length of $\sim59$ s and binned the Fourier frequency to an interval of $0.3$ Hz. The time lags as a function of frequency is shown in the
left panel of Figure~7. Here, the negative lags indicate the soft lags, which implies that the soft photons are lagging behind the hard ones. We have detected soft lags of $23.2\pm12.2$ ms at QPO frequency as marked with the dotted vertical line in the left panel of Figure~7. Furthermore, to investigate the nature of variation of time lags with energy at the QPO frequency, we considered $3-5$, $5-7$, $7-10$, $10-15$, $15-20$, $20-25$, $25-30$, $30-35$, $35-40$ and $40-60$ keV bands to derive the lag-energy spectrum. Here, the $3-5$ keV band was taken as the reference band and the Fourier frequency was binned to an interval of $0.3$ Hz. The time lags as a function of energy is shown in the right panel of Figure~7, which indicates the presence of soft lags (negative lags) in the $\sim6-7$ keV and above $\sim23$ keV bands. However, it is worth mentioning here that the soft lag (negative lag) detected in the  5-7 keV band is found to be statistically significant with the t-value of 2.31, whereas those estimated in the energy bands above $\sim23$ keV are statistically insignificant due to poor S/N ratio.  

Since, comparison of the time lags between the two consecutive failed outbursts in 2017 and 2018 can shed light on the nature of variability, we have performed the frequency resolved lag analysis using Stingray \citep{Huppenkothen2019a, Huppenkothen2019b} on data--4 taken by the XMM-Newton/EPIC-pn. In order to study the frequency-dependent time lag, we extracted two light curves in the $0.5-0.7$ keV and $0.7-5$ keV bands with time bin size of 0.5 s, which corresponds to the Nyquist frequency of 1 Hz. Each of these light curves were then divided into 3275 segments each having the length of 40 s, and the Fourier frequency was binned to an interval of $\sim0.3$ Hz. Left panel of Figure~8 shows the lag-frequency spectrum, where the soft lags of $140\pm80$ ms at the QPO frequency \citep[$\sim0.2$ Hz:][]{Stiele and Kong2021} is observed and depicted by the dotted vertical line. Furthermore, we also derived the lag-energy spectrum between a reference band and a series of adjacent energy bins using Stingray to investigate the dependence of time lags on energy at the QPO frequency. Here, we considered the lowest energy band as the reference band and averaged the time lags in the QPO frequency range. The right panel of Figure~8 shows the lag-energy spectrum, where the presence of soft lags has been observed at the energy band dominated by the iron emission line $(\sim6-8$ keV). 

\begin{figure*}
\plottwo{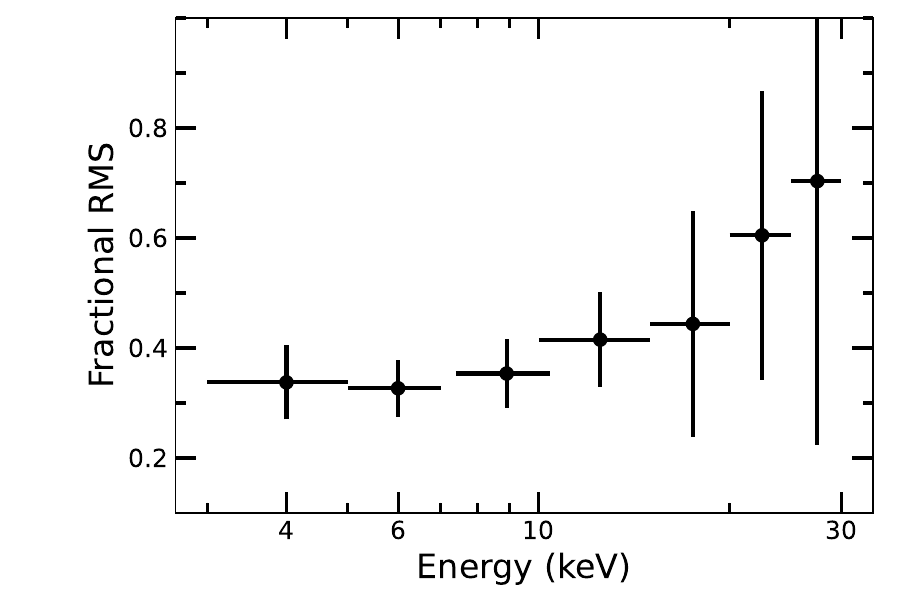}{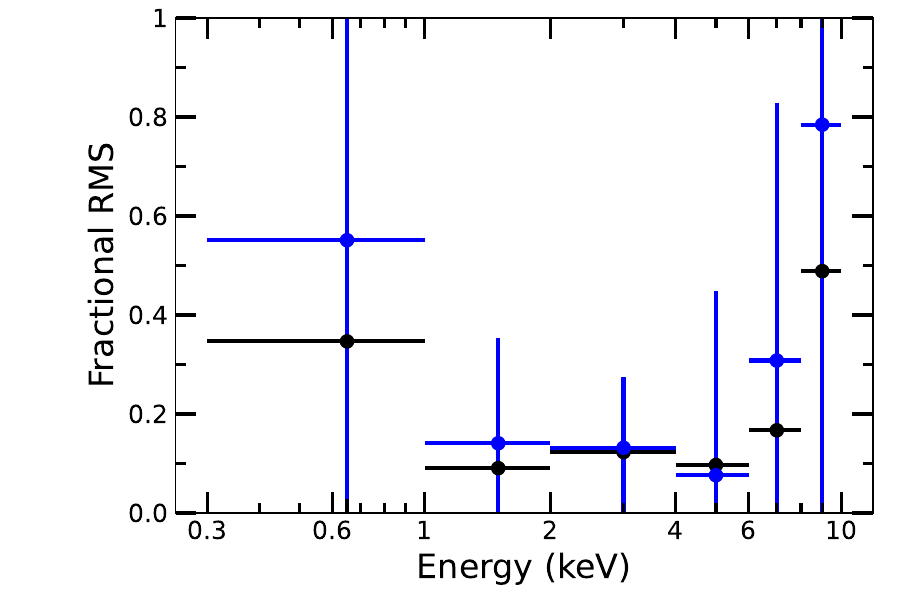}
\caption{Left panel: The fractional rms vs. energy for the frequency range of $0.1-10$ Hz derived using data--1 taken by AstroSat/LAXPC20.
Right panel: The fractional rms vs. energy for the frequency range of $0.1-10$ Hz derived using data--2 (black) and data--3 (blue) taken by Swift/XRT.}
\end{figure*}

\begin{figure}
\centering
\includegraphics[width=\columnwidth]{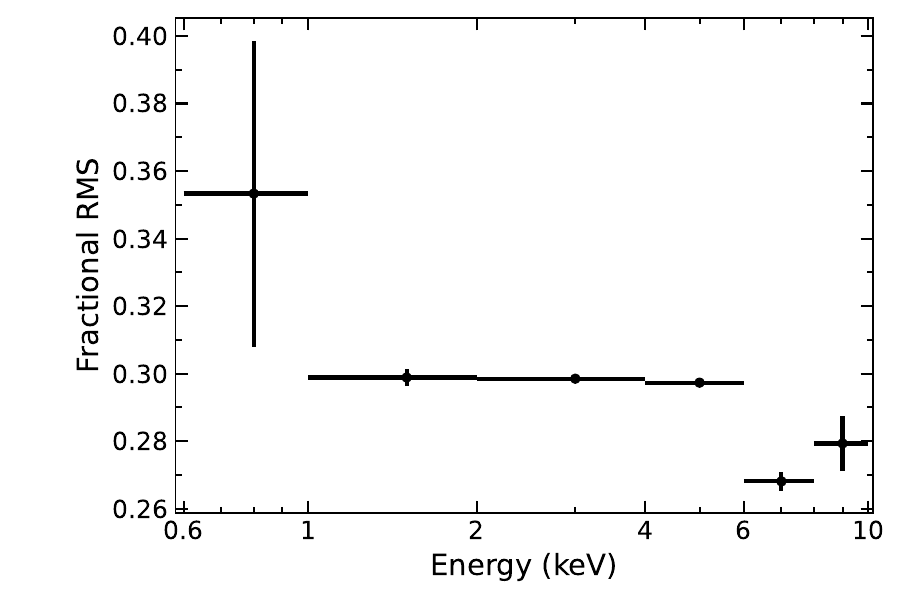}
\caption{The fractional rms vs. energy for the frequency range of $0.1-10$ Hz derived using data--4 taken by XMM-Newton.}
\label{fig:xmm-newton}
\end{figure}

\section{Broadband Spectral Analysis} \label{sec:floats}

 We fitted the SXT ($0.7-6$ keV) and LAXPC20 ($4-50$ keV) spectral data simultaneously to carry out the broadband X-ray spectral analysis in  the $0.7-50$ keV band using XSPEC \citep[V.12.10;][]{Arnaud1996}. The errors on the best fit parameters are quoted at $90\%$ confidence level. Using the \texttt{grppha} tool, we grouped the SXT and LAXPC20 spectral data into a minimum of 20 counts per bin. Initially, we fitted both the spectra jointly with a \texttt{powerlaw} model modified by the Galactic absorption component \texttt{TBabs}. Here, the abundances and cross 
 sections of the interstellar medium are considered from \citet{Wilms2000} and \citet{Verner1996}, respectively. In order to account for any difference in the relative normalizations between the two different instruments, we also multiplied a \texttt{constant} factor to the absorbed powerlaw model. The constant factor was fixed to 1 for the LAXPC20 spectral data and kept it free to vary for the SXT spectral data. Moreover, we incorporated a systematic error of $3\%$ to both the SXT and LAXPC20 spectral data to take care of the calibration uncertainty\footnote{\url{https://www.tifr.res.in/~astrosat_sxt/dataanalysis.html}}. In addition to this, a multicolor disk blackbody model \citep[\texttt{diskbb;}][]{Mitsuda1984} was further included to take care of the thermal emission from the accretion disk. As shown in Figure~9, the model setup \texttt{constant*tbabs(diskbb+powerlaw)} provided very poor fit and depicted the presence of iron line and the reflection hump through the prominent residual in the $5-9$ and above $\sim12$ keV bands, respectively. In this model setup, we then added a \texttt{gaussian} component to account for the iron line and replaced the \texttt{powerlaw} model with \texttt{nthcomp}, which describes the Comptonized emission coming from the hot coronal plasma \citep{Zdziarski et al.1996}. For the  \texttt{gaussian} component, we fixed the line energy at 6.5 keV and width at 0.1 keV considering a narrow line profile in the hard state and low spectral resolution 
of the LAXPC20 detector. An additional peak at $\sim30$ keV in the residual of LAXPC20 detector was also noticed, which is most likely due to the Xenon K emission feature \citep[see][]{Sridhar et al.2019}. In order to avoid any possible effect of this feature on the spectral parameters, we excluded the $25-31$ keV band while performing model fitting. As per the best-fit values of absorption column density available in literature, we fixed its value to be at $N_{\mathrm{H}} =
2\times10^{22}$ cm$^{-2}$ \citep{Parmar et al.2003, Miller et al.2006, McClintock et al.2009, Shidatsu et al.2014, Stiele and Yu 2016, Chand et al.2020, Chand et al.2021}. In addition to this, the seed photon temperature of the \texttt{nthcomp} model component was tied with the inner disk temperature of the \texttt{diskbb} model. As a result, this model setup \texttt{constant*tbabs(diskbb+gaussian+nthcomp)}(hereafter Model-1) provides an acceptable fit with $\chi^2/$dof$=542.9/541$. The best-fit spectral parameters are listed in Table~3, whereas the best-fit model along with spectra are shown in Figure~10. We found the best-fit photon index ($\Gamma$) to be $\sim1.65$, whereas the temperature of the inner accretion disk kT$_{in}$ is found to be $\sim0.24$ keV. The inner disk radius ($R_{\mathrm{in}}$) is derived from the \texttt{diskbb} normalization using the formula disk$_{\mathrm{norm}}$ $=(R_{\mathrm{in}}/D_{10})^2~\cos i$, where $R_{\mathrm{in}}$ is in the unit of km, $D_{10}$ is the distance to the source in the unit of 10 kpc, and $i$ is the inclination angle. We find $R_{\mathrm{in}}$ to be $\sim168$ km by assuming the distance to the source and the inclination angle ($i$) to be $\sim8.5$ kpc and $\sim75^\circ$, respectively \citep{Steiner et al.2012}. The electron temperature is found to be $kT_{\rm{e}} \sim13$ keV. Apart from this, the total unabsorbed flux is found to be dominated by the Comptonized emission, as the Comptonized fraction is found to be greater than $90\%$.

\begin{figure*}
\plottwo{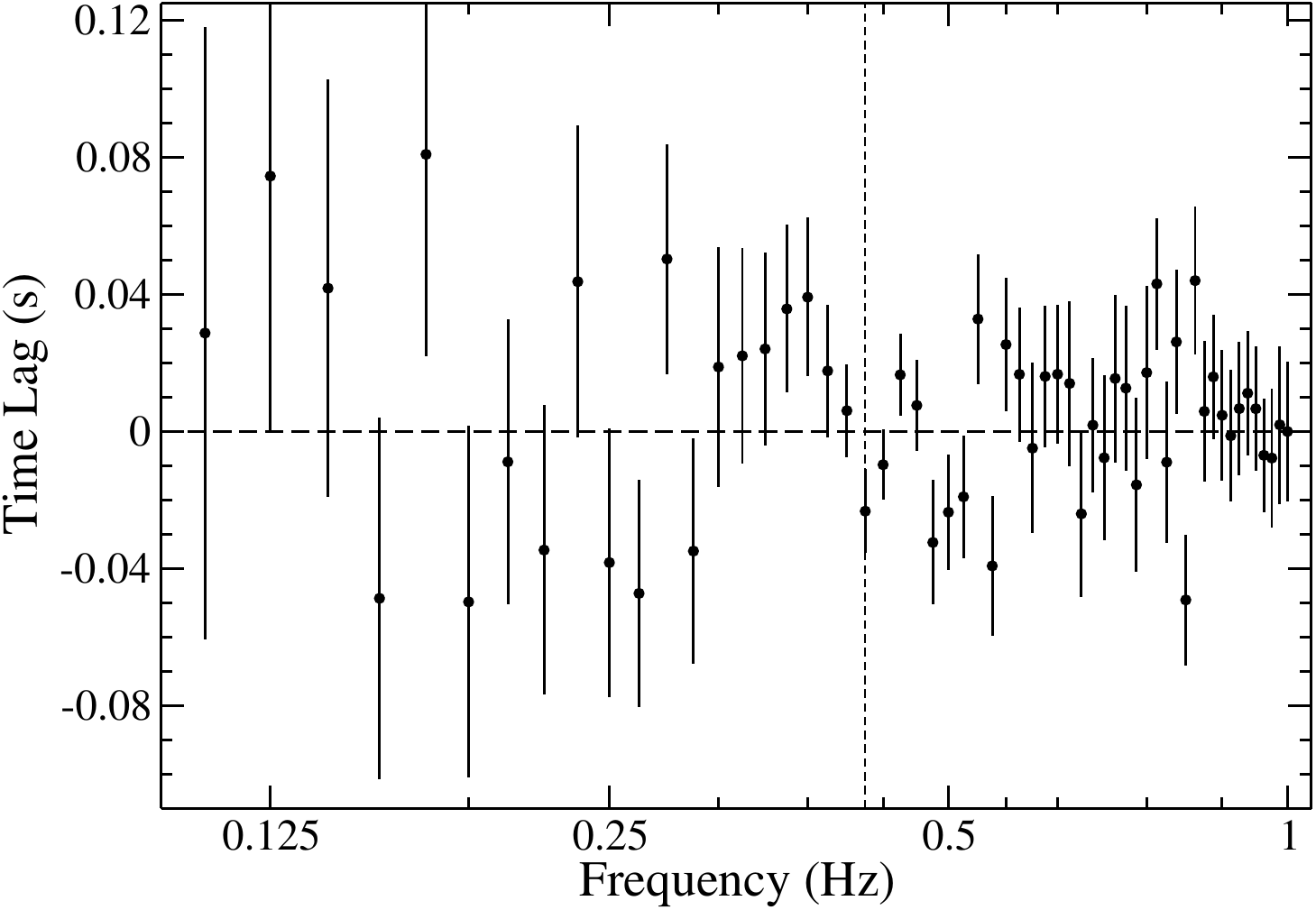}{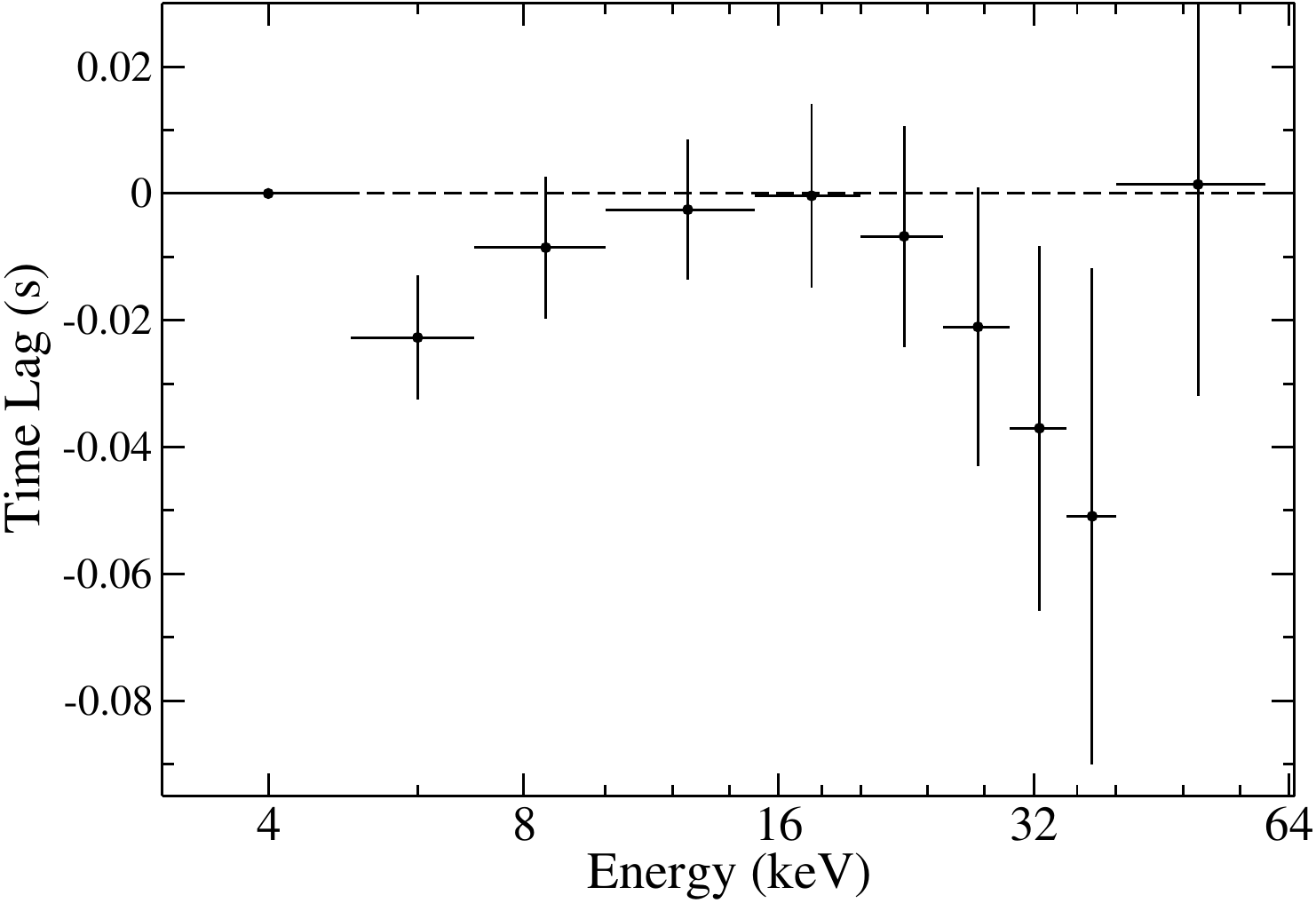}
\caption{Left panel: the frequency-dependent time lags between the $3-5$ and $9-13$ keV bands for the data--1 taken by AstroSat/LAXPC20. Vertical dotted line shows the position of the QPO, where soft or negative lag has been detected. Right pannel: the time lags as a function of energy derived at the QPO frequency ($\sim0.43$ Hz) using AstroSat/LAXPC20 observation.}
\end{figure*}

\begin{figure*}
\plottwo{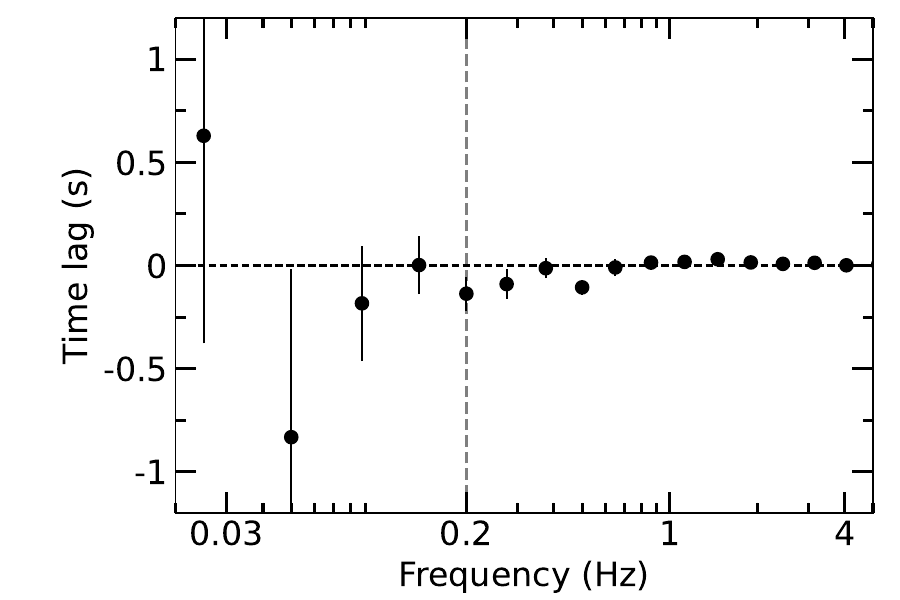}{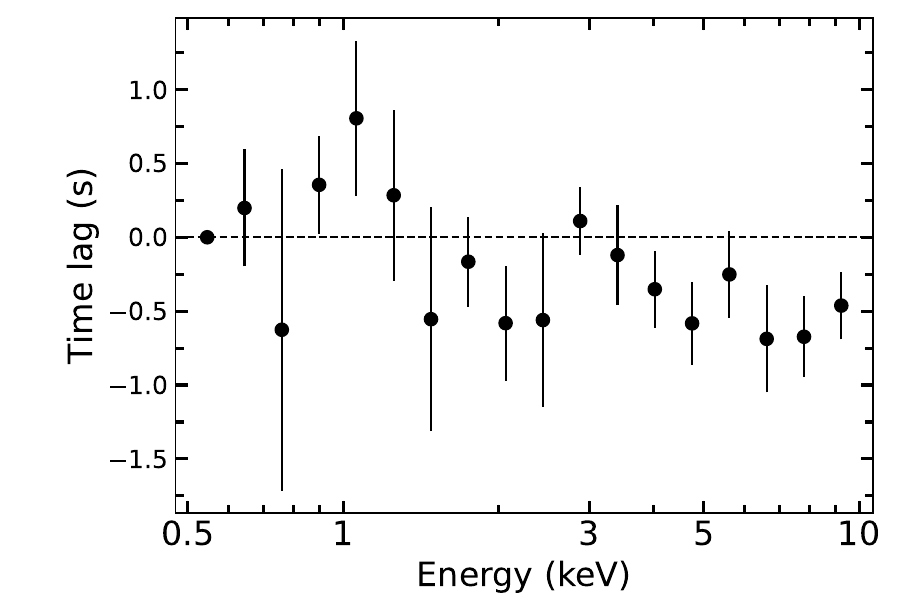}
\caption{Left panel: the frequency-dependent time lags computed between $0.5-0.7$ and $0.7-5$ keV bands for the data--4 taken by 
XMM-Neton/EPIC-pn. Vertical dotted line depicts the position of the QPO, where soft or negative lag has been detected. Right panel: the time lags as a function of energy derived at the QPO frequency ($\sim0.2$ Hz) using XMM-Newton/EPIC-pn observation.}
\end{figure*}

In the next scenario, we characterize the full reflection spectrum using a relativistic reflection model \texttt{relxillCp} from the \texttt{relxill} family \citep{Dauser2014, Garcia2014} in the $0.7-50$ keV energy band. This model \texttt{relxillCp} internally considers an incident spectrum in the form of an \texttt{nthcomp} Comptonization continuum. In this context, we replaced the
\texttt{gaussian} component in Model–1 and used \texttt{relxillCp} as a reflection component only by fixing the reflection fraction at $-1$. We tied the common parameters such as photon index ($\Gamma$) and electron temperature ($kT_{\rm{e}}$) between \texttt{nthcomp} and \texttt{relxillCp} components. In addition, spin of the black hole and disk inclination angle of the system were fixed at  $0.2$ and $75^\circ$, respectively \citep{Steiner et al.2012}. During the spectral fitting, we considered a single emissivity profile ($\epsilon \varpropto r^{-q}$, where $q$ is the emissivity index) over the whole accretion disk, and hence tied the break radius with the outer disk
radius at $400~r\rm{_g}$. Since the values of $q$ was pegging to the lowest defined value of 3, we fixed it at this value. We also fixed the iron abundance ($A_{\mathrm{Fe}}$) at the Solar value. Other parameters of the relxillCp component such as inner disk radius ($R_{\rm{in}}$), log of disk ionization ($log\xi$) and normalization were kept free to vary. 
Similar to Model-1, we also excluded the $25-31$ keV band to avoid the possible effect of the Xenon K emission feature on the best-fit spectral parameters. Thus, the model setup
\texttt{constant*tbabs(diskbb+nthcomp+relxillCp)} (hereafter Model-2) provided a reasonably better fit with $\chi^2/$dof$= 525.5/539$ in comparison to Model-1.
 All the best fit spectral parameters obtained from this model are listed in Table 3, whereas the best-fit model along with spectra are shown in Figure~10.  
It is worth mentioning here that the slight, narrow excess at $\sim0.9$ keV, seen in the residuals of Figure~10, is barely significant and appearing due to calibration uncertainties of the SXT. This feature does not also have any effect on the best-fit spectral parameters. We fixed $N_{\mathrm{H}}$ at the best-fit value of $2.2\times10^{22}~\rm{cm^{-2}}$, which is in well agreement with the previously reported values. We find the value of $kT_{in}$ to be the same within error as that obtained from Model-1. The inferred value of inner disk radius from the \texttt{diskbb} normalization using the formula stated above is $\sim334$ km. Although the inner  disk radius ($R_{\mathrm{in}}$) obtained from the reflection modeling could not be constrained, the lowest bound in the $90\%$ confidence limit is found to be $27.4$ $r_{\mathrm{g}}$. The values of $\Gamma$ and $kT_{\mathrm{e}}$ appear to be well consistent within error with those obtained from the Model-1. Furthermore, we find the  $log\xi$ to be $\sim3.6$. The unabsored flux derived in the $0.7-50$ keV from Model-2 remains same as that obtained from Model-1.

\begin{figure}
\centering
\includegraphics[width=\columnwidth]{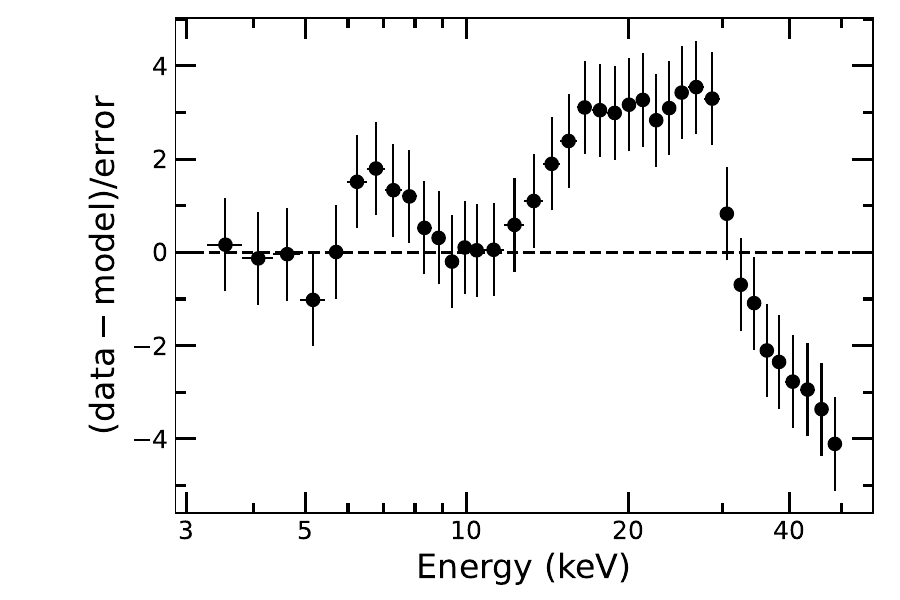}
\caption{Residual shows the deviations of the observed spectral data from the best-fit model {\tt tbabs(powerlaw+diskbb)}. The presence of the iron line excess in the $5-9$ keV region and the reflection hump peaking around $\sim15-30$ keV  band are clearly visible. Here, we have used only the LAXPC20 spectral data, and the spectral fitting is done only in $3-5$ and $9-12$ keV bands.}
\label{fig:xmm-newton}
\end{figure}

\begin{figure*}
\plottwo{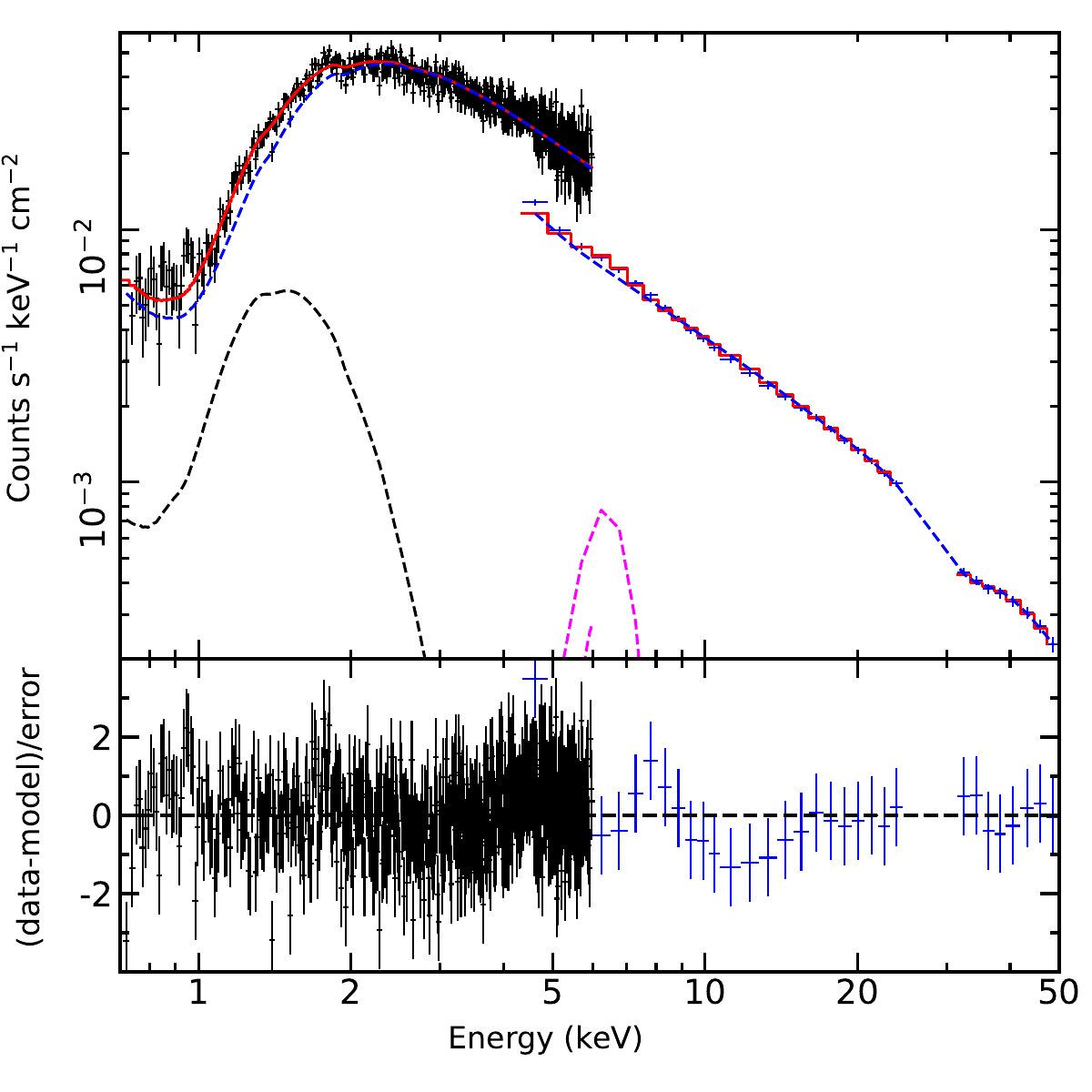}{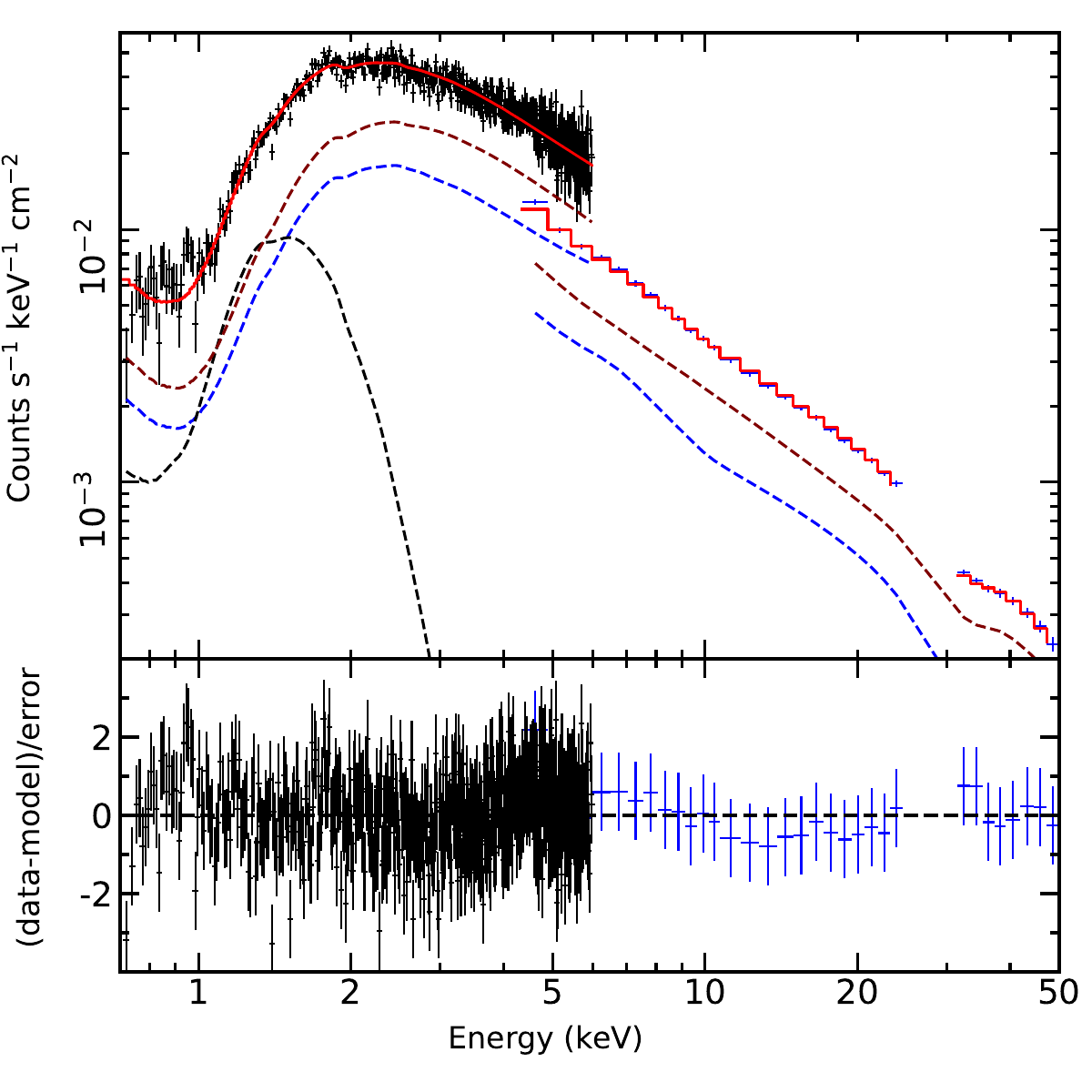}
\caption{Joint SXT and LAXPC20 counts spectra fitted with Model-1 (left panel) and Model-2 (right panel). Black and blue points indicate the SXT and LAXPC20 spectral data, respectively. The solid lines indicate the extrapolated models of the continuum fit. The dashed lines with different colors
indicate different model components: black (\texttt{diskbb}), pink (\texttt{gaussian}), blue (\texttt{nthcomp}), maroon (\texttt{relxillCp}).}
\end{figure*}

\begin{table*}
	\caption{Best-fit broadband X-ray spectral parameters derived using Model-1 and 2.}
   \centering
   \begin{tabular}{lccc}
      \hline
      \hline
	Component  & Parameter & Model-1 & Model-2 \\
	\hline
	
	TBabs & $N_H^1(\times 10^{22}$ cm$^{-2}$) & 2 (f) &  2.2 (f) \\ \\
 
    DISKBB & $kT_{in}^2$ (keV) & $0.24\pm0.02$ & $0.22\pm0.01$ \\
     & $R_{in}^{3}$ (km) & $168_{-44.6}^{+67.9}$ & $334^{+143}_{-93}$ \\ \\
    
    NTHCOMP & $\Gamma^4$ & $1.65\pm0.01$ & $1.66\pm0.02$ \\
    & $kTe^5 (keV)$ & $13.3_{-0.8}^{+1.0}$ & $17.1^{+2.5}_{-2.2}$ \\
    &$norm^6$ & $0.20\pm0.01$ & $0.13^{+0.05}_{-0.04}$ \\ \\
    
   GAUSSIAN & E$_{line}^7 (keV)$ & $6.5^f$ & ..\\
     & $\sigma^8 (keV)$ & $0.1^f$ &..\\
    &${norm}^9(\times10^{-3})$ & $1.9\pm0.8$ &..\\ \\
    
    RELXILLCp & $q^{10}$ & .. & $3^f$ \\
    & $a^{11}$ & ... & $0.2^f$ \\
    & $i^{12}$ ($^\circ$) & ... & $75^f$ \\
     &  $R_{in}^{13}$ ($r_{\mathrm{g}}$) & ... & $>27.4$ \\
    &  $R_{out}^{14}$($r_{\mathrm{g}}$) & ... & $400^f$ \\ 
    & $log\xi^{15}$ & ... & $3.6\pm0.3$ \\
    & A$_{\rm{Fe}}^{16}$ &... & $1^f$ \\
    &$\Re^{17}$ &... & $-1^f$\\
   &$norm^{18}$ $(\times 10^{-4})$ &... & $6.6^{+3.1}_{-3.3}$ \\ \\
   
   Flux$_{\rm{unabs}}^{19}$ $[10^{-9} \mathrm{erg~cm}^{-2} \mathrm{s}^{-1}]$  & & $2.7$ & $2.7$ \\
      & $\chi^2$/dof & $542.9/541$  & $525.5/539$ \\
	\hline 
    \end{tabular}

 \begin{tablenotes}
    \item \textbf{Notes.} $^1$Galactic absorption column density, $^2$ Inner disk temperature,$^3$  Disk radius from \texttt{diskbb} normalization, $^4$Photon index, $^5$Electron temperature, $^{7}$\texttt{gaussian} line energy, $^{8}$\texttt{gaussian} line width, $^{10}$Emissivity index,  $^{11}$Spin of the black hole, $^{12}$Disk inclination, $^{13}$Inner disk radius in the unit of $r_{\rm{g}}$, $^{14}$Outer disk radius in the unit of $r_{\rm{g}}$, $^{15}$Log of ionization parameter of the accretion disk, $^{16}$Iron abundance in the unit of Solar, $^{17}$Reflection fraction, $^{6, 9, 18}$ normalization parameter of the corresponding spectral components, $^{19}$Unabsorbed flux is derived in the $0.7-50$ keV band, f indicates the fixed parameters.
    \end{tablenotes}
\end{table*}

\section{Discussion and Conclusions}
In this paper, we have carried out a comprehensive temporal and broadband spectral analysis of the low-mass BHXRB H~1743--322 using the AstroSat and Swift/XRT observations taken during the 2017 outburst. Additionally, we have derived the time lags using XMM-Newton observation taken during 2018 outburst to compare its nature with that computed using AstroSat observation of 2017 outburst. From the HIDs (see Figure 1) derived using the Swift/XRT observations, we show that the outburst in 2017 of H 1743–322 was a failed one unlike its prior successful outburst in 2016 as reported by our group \citep{Chand et al.2020}. The outbursts of the source in 2017 and 2018 were also reported to be failed once by \citet{Stiele and Kong2021}.

As can be seen in Table 3, the best-fit photon indices ($\Gamma$) obtained from both the models suggest that the source was in the LHS during the  Astrosat observation of 2017 outburst. This is in well agreement with the position of the source on the HID.
Besides, the obtained value of inner disk temperature of $kT_{\rm{in}}$ $\sim0.2$ keV also supports the LHS scenario of the source, as well as appears to be the same as those  obtained during 2014 and 2018 failed outbursts \citep{Stiele and Kong2017, Stiele and Kong2021}. The dominance of the Comptonized emission over the total unabsorbed flux also signifies that the source might be in the LHS during this particular observation (see Section 4). The coronal electron temperature (kT$_\mathrm{e}\sim13-20$ keV) is also found to be quite close to that obtained during the 2014 failed outburst \citep [see][] {Stiele and Yu 2016}.

The inner accretion disk radius ($R_{\rm{in}}$) estimated from modeling either disk continuum or reflection spectrum, clearly indicates the presence of a truncated accretion disk (see Table 3). It is worth mentioning here that our findings of $R_{\rm{in}}$ from both the disk continuum and the reflection spectrum are consistent, whereas a few number of studies of BHXRBs have shown discrepancy in the inner extent of accretion disk in the LHS using these two methods \citep [see][] {Stiele and Kong2017, Garcia et al.2019, Connors et al.2022}. It is to be noted here that the extent of the inner accretion disk in the LHS is still a debatable topic. Through an ample amount studies, it is widely accepted that a truncated inner disk is introduced in the LHS, as a hot advection-dominated accretion flow (ADAF), replaces the geometrically thin and optically thick accretion disk whose emission dominates the X-ray continuum in HSS \citep[see][]{McClintock et al.1995, McClintock et al.2001, McClintock et al.2003, Narayan and Yi1995, Narayan et al.1996, Esin et al.2001, Ingram et al.2009, Furst et al.2015, Plant et al.2015, Ingram et al.2016}.
However, the inner extent of the accretion disk is also found in LHS by several studies \citep{Reis et al.2008, Miller et al.2010}, which were later on reported as an artifact due to the incorrect estimation of the pile up present in the data \citep{Yamada et al.2009, Done and Trigo2010, Kolehmainen et al.2014}. Later, \citet{Garcia et al.2019} studied the low mass BHXRB GX 339-4 and suggested by modelling the reflection component in the LHS that the inner accretion disk must extend close to the ISCO when the source luminosity is as high as $1\%$ of Eddington luminosity  ($L_{\rm{Edd}}$). In contrary to this result, the existence of a truncated disk at the luminosity greater than $5\%$ of Eddington luminosity was reported  by modeling the reflection spectrum for the low mass BHXRB MAXI J1820+070 by \citet{Zdziarski et al.2021}. Presence of truncated accretion disk in the LHS of this source is also reported by 
\citet{Banerjee et al.2024} using AstroSat observations. Similarly, we also find for 2017 failed outburst of H~1743--322 that the accretion disk is truncated away from the ISCO with $R_{\rm{in}}>27.4~r_{\rm{g}}$ at the source luminosity of $1.6\%~L_{\rm{Edd}}$, computed assuming the mass of the black hole to be $\sim11.2~M_\odot$ \citep{Molla et al.2017}. Moreover, considering the truncated disk picture of accretion flow in the LHS, a simple model was able to describe the fractional rms and lag-energy spectra at the QPO frequency of H~1743--322 during the 2017 outburst \citep [see][]{Husain2023}.

In the PDSs, we have detected LFQPO at $\sim0.43$ Hz derived from both data--1 and data--3. The presence of LFQPO is also observed in the PDS derived from data--2, however, at a comparatively lower frequency of $\sim0.35$ Hz. Apart from this, an upper-harmonic at $\sim0.9$ Hz, almost at the twice of the QPO frequency, is detected in the PDS from data--1. On the other hand, the upper-harmonic could not be detected in the PDSs from data--2 and 3, most likely due to their short exposures and the poor S/N ratio of the data. The shape of the PDSs and parameters of QPO such as frequency, Q-factor and fractional rms amplitude suggest that the observed QPOs from all the observations are type C in nature \citep{Casella et al.2004, Casella Belloni and Stella2005, Motta et al.2011}. It is worth to be noted here that the presence of type C QPO is very common in the LHS and HIMS \citep{Belloni2010}. As \cite{Husain2023} also reported the detection of a type C QPO along with upper harmonic in a single energy band of $3-15$ keV using the AstroSat observation, we confirm their detection and further demonstrate the energy independent nature of these features over the three different energy bands (see Table 2). In addition, it is also clear from Table 2 that over the period of three days of Swift/XRT observations, the QPO frequency shifts by $0.08\pm0.02$ Hz towards the higher frequency side as the outburst evolves. An increase in the QPO frequency with time has also been noticed for H~1743--322 \citep{Chand et al.2021} and an another BHXRB XTE~J1550--564 \citep{Dutta and Chakrabarti2016} during the evolution of 2016 and 1998 outbursts, respectively. Although the exact origin of the QPO is still unknown, speculations from different studies suggest that its origin is associated  with either corona \citep{Titarchuk and Fiorito2004} or Lense–Thirring precession of a radially extended region of the hot inner ﬂow in the truncated disk model \citep{Stella and Vietri1998, Ingram et al.2009}. The increase in the QPO frequency at the rising phase of an outburst is suggested to be associated
with the decrease in coronal size by \citet{Chakraborty et al.2008,  Chakraborty et al.2009} and \citet{Dutta and Chakrabarti2016}. Moreover, \citet{Zhang et al.2022} reported that the change in the type C QPO frequency is related to the evolution in the morphology of the corona.  

As can be seen from the left panel of Figure 5, the fractional rms variability amplitude in the $0.01-10$ Hz for data-1 taken by AstroSat/LAXPC20, remains almost stable up to $\sim10$ keV, and then increases at higher energies. Despite the larger error bars, the Swift/XRT observations also show an increasing trend in the rms-energy spectra at $\gtrsim 6$ keV band (see the right panel of Figure 5). Besides, the increasing trend at $\lesssim 1$ keV band may give hint about
the presence of some disk variability in the source on longer timescales. On the other hand, the increasing trend in the rms-energy spectra at higher energies is being triggered by Comptonized and reflection components as we have noticed from the broadband spectral analysis that the disk component contributes only up to $\approx 3$ keV (see Figure~10). Apart from this, the possible disk variability can also be noticed from the rms-energy spectrum derived using the XMM-Newton
observation during the declining phase of 2018 outburst (see Figure~6), where a significantly large variability is observed at $\lesssim 1$ keV. In this regard it is worth to mentioning here that, the covariance spectra, derived by \citet{Stiele and Kong2021} using Nicer observations during the early phase of 2018 outburst, depict large variability at $\lesssim 1$ keV (see their Figure~7), which in turn may imply the disk variability.

The  presence of negative or soft lags ($23.2\pm12.2$ ms) at the QPO frequency of 0.43 Hz during 2017 outburst, indicates the soft photons in the $3-5$ keV band are delayed with respect to the hard photons in the $9-13$ keV band (see Figure~7). In addition to this, the trend of hard time lags as seen at the relatively lower frequency ($\sim0.35-0.4$ Hz) is often observed in the BHTs in the LHS as low frequencies (long timescales) are usually dominated by the presence of hard lags \citep [cf.][]{Papadakis et al.2001, DeMarco et al.2015}. On the other hand, the soft time lags are typically observed at high frequencies (short timescales) as the hard lags are suppressed at these frequencies \citep [cf.][]{DeMarco et al.2015}. Presence of soft time lags at the QPO frequency in the AstroSat observation during 2017 outburst is also reported by \citet{Husain2023} although they considered different energy bands. In comparison to the AstroSat observation, we find that the soft lag amplitude ($140\pm80$ ms) at the QPO frequency of $0.2$ Hz for the XMM-Newton observation during 2018 outburst is significantly larger (see Figure~8). We note that assumption of different energy bands during time lags calculation may also have an effect on the lag amplitude. The outburst in 2018 of H~1743--322 has a similarity with its 2017 outburst in the sense that both are failed in nature. As similar to our results using  the AstroSat observation during 2017 outburt, the source was found to be in the LHS and the accretion disk was significantly truncated away from the ISCO for the XMM-Newton observation during 2018 outburst \citep [see][]{Stiele and Kong2021}. 

Apart from this, the signature of soft lags in the frequency resolved lag-energy spectrum at $\sim6-7$ keV for the AstroSat (data--1) and XMM-Newton (data--4) observations, as well as above the 23 keV in the AstroSat (data--1) observations can be associated with the reflection phenomena as these bands are usually dominated by iron emission line and reflection hump, respectively. Although the detected soft lags above $\sim 23$ keV is statistically insignificant due to poor S/N ratio, it still provides a hint on its association with the reflection hump region. As an early suggestion by \citet{Reig et al.2000}, the origin of the soft and hard time lags are related to the Compton up and down-scattering of photons in the coronal plasma, which may have different temperature gradient and optical depth. However, in the case of AGNs, the responsible phenomena behind both the hard and soft lags is thought to be the reverberation, arising due to the light travel time delay between the primary X-ray continuum and reprocessed emissions from the accretion disk \citep{Zoghbi et al.2010, Wilkins et al.2013}. Recently, the reverberation time lags, originating due to thermal reverberation, are also detected in the BHXRBs such as GX~339--4 and H~1743--322 \citep{DeMarco et al.2015}. Using the XMM-Newton observation of H 1743-322 during 2014 failed outburst, \citet{DeMarco and Ponti2016} detected a soft lags of $\sim60$ ms and suggested the origin of the soft lags to be thermal reverberation. Furthermore, using the RXTE observations of GX~339--4 and XTE~J1550--564, \citet{Dutta and Chakrabarti2016} suggested that the soft time lags at the QPO frequency may appear in the highly inclined ($>60^\circ$) sources as a combined effect of Comptonization, reflection and focusing effect. On the other hand, the reflection and focusing effects become negligible for the sources with low inclination angle ($<60^\circ$). Dependency of time lags at type C QPO frequency on the orbital inclination angle is also reported by \citet{van den Eijnden et al.2017}. In order to investigate further the origin of the soft lags and their potential association with reverberation phenomena in the hard-only or failed outbursts, high-quality observations will be required, which may also shed light on the accretion mechanism of BHTs during such failed outbursts.

\section{acknowledgments}
We thank the anonymous referee for the useful comments and suggestions that have improved the quality of the paper.
The authors acknowledge the financial support of ISRO under the AstroSat archival Data utilization program (No: DS-2B-13013(2)/8/2019-Sec.2).
This publication uses data from the AstroSat mission of the Indian Space Research Organisation (ISRO), archived at the Indian 
Space Science Data Centre (ISSDC). This work has used data from the SXT and LAXPC instruments on board AstroSat. The LAXPC data were
processed by the Payload Operation Center (POC) at TIFR, Mumbai. This work has been performed utilizing the calibration databases and auxiliary analysis tools developed, maintained, and distributed by the AstroSat-SXT team with members from various institutions in India and abroad and the SXT POC at the TIFR, Mumbai ({\url{https://www.tifr.res.in/~astrosat_sxt/index.html}). The SXT data were processed and veriﬁed by the SXT POC. This research has also made use of data supplied by the UK Swift Science Data Centre at the University of Leicester and MAXI data provided by RIKEN, JAXA, and the MAXI team.The Science Analysis System (SAS), provided by the ESA science mission with instruments and contributions directly funded by ESA Member States and the NASA (USA) is used to process the
XMM-Newton data. P. S. acknowledges Guru Ghasidas Vishwavidyalaya (A Central University), Bilapur(C.G.), India,  for provding fellowship under VRET scheme. P. T. expresses his sincere thanks to the Inter-University Centre for Astronomy and Astrophysics (IUCAA), Pune, India, for granting support through the IUCAA associateship program. V.K.A. thanks GH, SAG; DD, PDMSA and Director, URSC for encouragement and continuous support to carry out this research.
\vspace{5mm}
\facilities{AstroSat (SXT, LAXPC), Swift(XRT), XMM-Newton(EPIC-pn), MAXI.}

\software{XSPEC \citep{Arnaud1996}, Julia, LAXPCSoft, Matplotlib, Stingray.}



\end{document}